\documentclass[prd, preprint, groupedaddress, nofootinbib]{revtex4-2}
\pdfoutput=1
\usepackage[utf8]{inputenc} 
\usepackage{amssymb}
\usepackage{amsmath}
\usepackage{amsthm}   
\DeclareMathOperator{\diag}{diag}

\usepackage{braket}
\usepackage{bbm}
\usepackage{revsymb} %for lambda bar

\usepackage{graphicx}
\usepackage[caption=false]{subfig}

\usepackage{color}

\usepackage{soul}

\usepackage{siunitx}

\usepackage{letltxmacro}
\let\oldeqref\eqref
\RenewDocumentCommand\eqref{oom}{%
\IfNoValueTF{#2}{\def\eqafter{}}{\def\eqafter{#2}}%
\IfNoValueTF{#1}
{\oldeqref{#3}}
{(#1\textup{$\,$\ref{#3}$\,$}\eqafter)}%
}

\newcommand{\vect}[1]{\boldsymbol{#1}}

\usepackage[hidelinks]{hyperref}     
\hypersetup{colorlinks=true, linkcolor=black, linktocpage=true,  citecolor=blue}

\begin{document}

\title{Deep learning to detect gravitational waves from binary close encounters:\\ fast parameter estimation using normalizing flows}

\author{Federico De Santi}
\email[]{f.desanti@studenti.unipi.it}
\author{Massimiliano Razzano}
\email[]{massimiliano.razzano@unipi.it}
\author{Francesco Fidecaro}
\author{Luca Muccillo}
\author{Lucia Papalini}
\author{Barbara Patricelli}

\affiliation{Dipartimento di Fisica "Enrico Fermi", Universit\`a di Pisa, I-56127 Pisa, Italy\\ }
\affiliation{Istituto Nazionale di Fisica Nucleare, Sezione di Pisa, I-56127 Pisa, Italy\\ }
 
\date{April 2, 2024}
\newcommand{\barbara}[1]{{\em \color{magenta}{Barbara:} }{\color{magenta}{#1}}} 
\begin{abstract}
A yet undetected class of gravitational wave signals is represented by the  close encounters between compact objects in highly-eccentric ($e \sim 1$) orbits, that can occur in binary compact systems formed in dense environments such as globular clusters. The expected gravitational signals from these close encounters are short-duration pulses that would repeat over a much longer time scale in case of multiple  passages at periastron. These sources represent a unique opportunity of exploring astrophysical formation channels as well as a different way of testing general relativity. Furthermore, in the case of binary systems containing  neutron stars, the observation of these sources could help to constrain the neutron star equation of state, thanks to the signature left in the gravitational wave signal by the f-modes excitation that can occur during the encounter. 

The detection and parameter estimation of these signals is however challenging given the short duration of expected signals and the sensitivities of current ground-based gravitational wave interferometers.
We present a novel approach to perform fast detection and parameter estimation of gravitational wave signals from binary close encounters that exploits probabilistic machine learning. We have used Conditional Normalizing Flows to model complex probability distributions and therefore infer posterior distributions for the source parameters. This architecture is able to perform inference in a very short time and its output can be directly compared with classical methods. Fast detection and parameter estimation is very important as it could trigger electromagnetic follow-up campaigns and offer the possibility to study these events in a multimessenger context.
To develop and test the algorithm, we have focused on the simulations of single bursts emission obtained using the Effective Fly-by formalism and embedded in the noise of Advanced LIGO and Virgo during their third Observing Run (O3).
Our proposed model outperforms standard Bayesian methods in accuracy and is $\sim$5 orders of magnitude faster, being able to produce $5\times 10^4$ posterior samples in just \SI{0.5}{\,s}. The results are extremely promising and constitute the first successful attempt for a fast and complete parameter estimation of binary close encounters using deep learning, offering  a new approach to study the evolution of orbital parameters of compact binary systems.
\end{abstract}

\maketitle
\section{Introduction}\label{Introduction}

The detection of gravitational waves represents a revolution in the way we probe the Universe and provides a new and independent tool to investigate the physics of extreme compact objects.  For instance, the first detection of gravitational waves from the coalescence of a binary black hole system,  GW150914 \cite{GW150914}, provided the observational proof of the existence of stellar-mass black holes with masses greater than $\simeq 25 M_\odot$ and established that binary black holes can form in nature and can merge within a Hubble time. Furthermore, the detection of gravitational waves from the event GW170817 and its associated electromagnetic counterparts marked the birth of a new era in multimessenger astrophysics \cite{GW170817, Multimessenger_Observation_GW170817}. 
The joint observation of electromagnetic and gravitational waves provided the first confirmation that binary neutron star coalescence are progenitors of short Gamma-Ray Bursts \cite{2017ApJ...848L..13A}, 
and allowed the investigation of the origin of heavy elements \cite{2017Natur.551...67P,2017Natur.551...75S}. Furthermore, multimessenger observations of GW170817 offered a new way of investigating the equation of state of neutron stars \cite{Radice_2018_NS_EOS, MARGALIT_EOS}, testing General Relativity \cite{GR_tests_GW170817} and measuring the Hubble constant \cite{Nature_Hubble_constant}. \\
The third Gravitational Wave Transient Catalogue (GWTC-3) \cite{gwtc3} contains 90 events detected by Advanced LIGO and Virgo during the first three observing runs (O1, O2, O3) from 2015 to 2020. All these events are associated with the coalescence of compact binary systems (CBCs) containing black holes  and/or neutron stars. 
More specifically, several dozens are consistent with binary black hole (BBH) systems. The growing population of BBHs observed through gravitational waves allowed to perform population studies that seem to support the presence of more than one binary formation channel \cite{gwtc2-population, gwtc3_population}. There seem to be two main formation channels \cite{comparison_gw_formation_channels}: BBHs can be the outcome of isolated binary evolution, i.e. they can form from the evolution of stars paired together at birth, or they can form dynamically, through strong stellar encounters in dense environments as young, globular, and nuclear clusters or active galactic nuclei. A deeper understanding of these different formation mechanisms is crucial in order to fully explain the BBH population so far observed. \\
Recent simulations of dynamical interactions in globular clusters have predicted the existence of populations of binaries merging with non-null eccentricity ($e > 0.05$) \cite{Samsing_2018, Post-Newtoniandynamics_GC_general}. 
Despite gravitational wave emission being in general an efficient mechanism for the orbit circularization during the binary evolution, these works have revealed the existence of BBH subpopulations forming in orbits with eccentricities $e\sim 1$. We will refer to them as \text{Close Encounters} (CE). \\

Accurate measurement of the parameters of CE signals is of paramount importance to study dynamical formation channels as well as gravity in the strong field regime. At the moment no confident gravitational wave signal emitted during a CE has been detected, making these sources new and potentially interesting to search for \cite{hyperbolic_encounters_not_found, Morr_s_2022}.\\
Due to the high eccentricity of these systems, the expected gravitational wave emission differs from the chirp-like waveform detected from CBCs. Eccentricity induces a modulation in the waveform that, in the limit $e\rightarrow1$, transforms it into a series of repeated short duration bursts emitted during each periastron passage.\\
The burst-like nature of the signal, combined  with the expected low signal-to-noise ratio, makes the detection of these sources particularly challenging. While current search strategies are based on unmodeled searches, Deep Learning has been proposed as a possible new approach to analyze these sources \cite{CE_deep_learning, Gleetter_Nunzio}.\\
This paper present a novel approach to the detection and parameter estimation of gravitational waves from CEs based on Probabilistic Machine Learning. 
Our approach exploits Normalizing Flows (NFs) to combine Bayesian inference methods with Deep Learning. This approach has been successfully tested on other types of sources. For instance,  BBH coalescences have been studied with DINGO \cite{green2020gravitational,dax2021real}. 
We will focus on single burst emission from encounters of binary black hole systems, as they are ones most likely to be detected by the current generation of interferometric detectors. We defer to subsequent work the application to the case of repeated bursts.   
The paper is organized in this way. In Section \ref{CE_physics} we discuss the dynamical scenarios for the formation of CEs and their expected gravitational wave emission derived from the \textit{Effective Fly-by} formalism. Section \ref{sec 3: Normalizing Flows} introduces Normalizing Flows and their properties. In Section \ref{Description of Architecture} we discuss HYPERION, the NF-based pipeline that we have developed for parameter estimation using NFs. Section \ref{Simulations and results} contains the training on a simulated dataset and the resulting performance of the pipeline. Finally, section \ref{Discussion} discusses the results and limitations of this approach. 

\section{Binary Close Encounters as gravitational wave sources}\label{CE_physics}

The canonical formation channel for BBH systems is via isolated binary evolution driven by stellar physics \cite{Giacobbo_Mapelli_2018}. Stellar evolution further predicts the existence of a gap in the BH mass distribution from $50_{-10}^{+20}\; \si{M}_{\odot}$ to approximately $120\; \si{M}_{\odot}$ because of pair-instability supernovae. The main uncertainties in the boundaries of this mass gap  are related to limited knowledge of processes at play during the evolution of massive stars: e.g. the $^{12}\si{C}(\alpha, \gamma)^{16}\si{O}$  reaction \cite{gwtc3_population}. 

However, population studies, made possible thanks to catalogs of observed gravitational wave events, have revealed a slightly different picture. In particular, the inferred distribution for the primary mass component in GWTC-3 
does not exhibit a sharp drop at $\sim 50 \;\si{M}_\odot$ \cite{gwtc3_population} as one would expect from the outlined formation channel.
The presence of a tail at higher masses seems to suggest that a fraction of the observed BBHs could have formed through additional formation channels that have to be of \text{dynamical} origin, i.e. from \text{N-body} interaction between stars and/or black holes. 

Besides the mass distribution, another ingredient that can provide clues to the formation channel is the spin orientation of the binaries. 
For instance, isolated field binary evolution is believed to produce components with preferably aligned spins \cite{aligned_spins_field_binaries} in contrast to dynamical encounters which can lead to isotropic spin-orbit misalignment \cite{spin_misalignment}. 
There are currently evidences for the spin distribution to require misalignment as well as events with anti-aligned spins \cite{gwtc3_population}; this could  suggest that some of the observed BBHs formed dynamically, but further investigations are needed. 

Therefore, this has led to the examination of these additional channels, which are possible in highly dense stellar environments. Examples of such environments are globular clusters which have central densities $\rho_c \geq 10^4\,\si{M}_\odot\, \si{pc}^{-3}$ \cite{Mapelli_2021}, \text{young stellar clusters} with $\rho_c > 10^3 \,\si{M}_\odot\, \si{pc}^{-3}$ \cite{Mapelli_2021, Mapelli_YMC}, nuclear star clusters of galactic nuclei \cite{eccentric_binaries_galactic_nuclei} as well as active galactic nuclei \cite{BBH_in_AGNs}.

\subsection{Dynamics in Dense Stellar Environments}\label{sec2: GC interactions}
Given the high stellar density in globular clusters, \text{single-single}, \text{binary-single} and even \text{binary-binary} interactions can take place and influence the evolution of binary systems.
These interactions have been studied through numerical N-body simulations and have revealed a wide spectrum of possible final states \cite{Zevin_2019_4-body&Zevin_2019_plot_populations, Samsing_2014}.\\
Recent simulations in globular clusters \cite{Samsing_2014, Samsing_2018} have indeed confirmed that multiple resonant interactions can lead to the formation of highly eccentric compact binaries. 
A subset of these binaries forms in a condition in which energy loss due to gravitational wave emission produces a \text{capture}: the inspiral phase of the binary speeds up, leading to a merger with a non-negligible eccentricity. More importantly, a subset of them is expected to merge within the \text{LIGO-Virgo-KAGRA} frequency range \cite{Post-Newtoniandynamics_GC_general}.

%___________________________________________________

\subsection{Highly Eccentric Compact Binaries in Globular Clusters: populations and rates}\label{subsec2: Pops and rates in GCs}

Dynamical interactions in globular clusters can produce different populations of merging systems, each one with its typical eccentricity. The dominant frequency $f_{GW}^{peak}$ at which these binaries emit gravitational waves is: \cite{Post-Newtoniandynamics_GC_general}

\begin{equation}\label{eq: peak GW freq}
    f_{GW}^{peak} = \frac{\sqrt{GM}}{\pi}\frac{(1+e)^{1.1954}}{[a(1-e^2)]^{3/2}}
\end{equation}
with $M$ being the total mass of the binary, $a$  its semi-major axis and $e$ the eccentricity .\\
BBHs mergers formed through dynamical interactions in globular clusters fall into three major categories \cite{Post-Newtoniandynamics_GC_general, Samsing_2014}, depending on the timescale $T_{GW}$ for gravitational wave emission to drive a binary to merge and the average timescale $T_{SE}$ between two successive encounters. In particular they are defined as \cite{Post-Newtoniandynamics_GC_general}:
\begin{equation}
    T_{GW} \propto a^4(1-e^2)^{7/2}\;, \quad T_{SE} \propto na^2\sigma \left( 1+ \frac{GM}{2a\sigma^2} \right)
\end{equation}
where $n$ and $\sigma$ are the number density and the velocity dispersion in the cluster, respectively. \\ 
The first category of BBH mergers is that of ejected inspirals, which are binary systems that, by the recoil from close interaction, acquire a center of mass velocity that exceeds the escape velocity of the cluster and get ejected from it. These mergers produce gravitational waves with $f_{GW}^{peak} \leq 10^{-2}\;\si{Hz}$ while being characterized by a non-zero eccentricity ($e>0.01$) \cite{Samsing_2018_LISA, Ejected_inspirals_LISA}. For this reason, they are among the major sources detectable by \text{LISA} \cite{LISA}.\\
The In-cluster mergers are a second category of binaries merging inside the cluster due to dynamical encounters, but \text{not} due to significant emission of gravitational waves during the encounters. They can be of two kinds: \text{2-body} and \text{3-body mergers}. The former are binary black holes that survive a binary-single interaction with semi-major-axis and eccentricities such that their inspiral times are less than interaction times ($\sim 10^7$ years \cite{Samsing_globular_cluster_pic}). Their eccentricity is expected to be similar to that of ejected inspirals and the $f_{GW}^{peak}$ near the \text{LISA} sensitivity band \cite{Samsing_LISA_II}. The latter are still formed through binary-single interactions. However, their pericenter distance is perturbed in such a way that the energy lost over one orbit through gravitational wave radiation is larger than the initial energy of the 3-body system. Timescales associated with this process are thus much smaller ($\sim1$ year), which implies gravitational waves frequency peaks in the ground-based detector sensitivity bands \cite{Samsing_globular_cluster_pic, Samsing_LISA_II}.\\
Finally, the category of gravitational wave captures consist of binaries that inspiral and merge during a resonant interaction itself due to the strong emission of gravitational waves. This interaction can be a binary-single, binary-binary, or even a single-single. In the latter, two initially unbound objects experience an encounter on a hyperbolic orbit that causes the binary to become bound and rapidly merge. They typically result in $f_{GW} \geq 10^{-1}\;\si{Hz}$. However, this mechanism is also able to produce highly eccentric binaries ($e \sim 1$) that will merge within the sensitivity band of ground-based detectors with timescales $\mathcal{O}$(seconds). Given the high eccentricities of this last subset, some of them are close enough to the unbound limit to experience \text{fly-by} encounters \cite{Post-Newtoniandynamics_GC_general}.
The expected rate of eccentric BBH captures is expected to be $1-2$\SI{}{\;\giga pc^{-3}\,yr^{-1}}
in the local universe ($z < 1$) \cite{Post-Newtoniandynamics_GC_general}.

%___________________________________________________

\subsection{Astrophysical Relevance of CE Observations}\label{subsec2: Astrophysical implications of CE}

Close Encounters carry distinctive signatures that can be used to differentiate between different formation channels, hence probing the underlying mechanisms responsible for the binary formation and merger. Tests of General Relativity can also be carried out with such sources. For eccentric bound orbits, the smallest pericenter distance can be $r_p/M \sim4$ ($G=c=1$ units) corresponding to $v_p \sim 0.7c$ \cite{Loutrel_2021}. Therefore, CEs provide themselves as a unique laboratory to test General Relativity in the \text{strong-field regime}: higher order effects such as radiation reaction and tides are indeed expected to become dominant. Other than that, eccentricity can be used to put constraints on alternative or modified theories of Gravity \cite{Loutrel_2014_GR_tests}.
\text{Neutron Star's Equation of State} can also be constrained if one of them is present in the binary. In the case of CBCs, the effects related to the equation of state become relevant only during the late inspiral and post-merger phase. In the case of eccentric inspiral, on the contrary, \text{f-modes} on the NS surface can be excited during each close interaction \cite{Loutrel_f_modes}.
CE events could be potentially interesting also from a multi-messenger point of view, either when neutron stars \cite{EM_counterparts_from_NS_CE} and/or BHs are involved. As already mentioned, CEs can happen between two BHs embedded in the accretion disk of an  active galactic nuclei \cite{BBH_in_AGNs} and, in such a gas rich environment, the merger can also yield a significant, detectable EM counterpart (see e.g. \cite{CE_multimessenger_I, CE_multimessenger_II}). 

CEs could also be the source of a stochastic background from primordial black holes. Close Hyperbolic Encounters from primordial black oles have been recently proposed as a detectable source for Einstein Telescope \cite{stochastic_CE}. Being not resolvable, this emission results in overlapping bursts forming a stochastic background. 
In this work we will consider BBH gravitational wave captures at high eccentricity ($e \sim 1$), since they are expected to be detectable with current ground-based interferometers.

\subsection{Waveforms for Eccentric Close Encounters}\label{sec2: Waveforms for Close Encounters}
In order to be able to infer the parameters of a CE source, an accurate theoretical description of the gravitational wave signal emitted is needed.
The presence of eccentricity, which is the defining feature of these system, poses several challenges. In first place, it makes mandatory to have accurate waveforms models. Indeed, even a small orbital eccentricity, if not correctly accounted for, is able to introduce systematic biases that exceed the statistical errors in parameter estimation \cite{eccentricity_importance_in_PE}. As an example in \cite{Guo_mimicking_mergers_and_captures} it has been shown that black hole captures might be misclassified as standard CBCs.\\
Currently, the most accurate gravitational wave waveforms are obtained through numerical relativity simulations, which have the drawback of being extremely computationally costly. This is due to the great velocities reached during the encounter which impose small integration steps. On the other hand, successive periastron passages happen on much wider timescales. Numerical relativity simulations available today, hence, only cover a limited number of orbits \cite{zoom_whirls_NR_simulations, NR_simulation_2} and have shown that the gravitational wave emission consists in a series of repeated \text{burst signals}. \\
Since numerical relativity waveforms are too expensive to be exploited during an online analysis, it is crucial to also pursue an analytical approach. In order to account for relativistic effects such as \text{radiation reaction}, the Post Newtonian formalism is widely used. This method, which works well for binaries in quasi-circular orbit, has difficulties in the high eccentricity limit since it is based on a \text{post-circular expansion} where the eccentric orbit is seen as a perturbation of a circular one. Previous attempts to describe eccentric waveforms in this way have been done in \cite{Post_circular_expansion, Post_circular_expansion_PN} up to eccentricities $\lesssim 0.8$ for widely separated binaries. However this approach suffers from Post Newtonian convergence issues when considering higher $e$ or smaller separations \cite{Loutrel_2021}. Therefore the description of Close Encounter mergers is not fully feasible with it.\\
An alternative solution is represented by a formalism recently developed: the Effective Fly-by formalism \cite{Loutrel_2020, Loutrel_2021}.
The difference with respect to other analytical approaches is that the periastron passage in the eccentric orbit is obtained by perturbing a \text{parabolic fly-by}. That defines the \text{post-parabolic approximation} \cite{Loutrel_2020}. Hence, the Effective Fly-by Formalism  provides an accurate analytical description of the single burst emission at each periastron passage by modeling the single close passage as \text{fly-by}: i.e., a perturbation on a parabolic orbit. This method overcomes the issues of the post-circular approximation and is best suited for higher eccentricities.
It is also possible to derive the whole inspiral waveforms from a single burst by adding many of these. In order to do so, it is necessary to include radiation reaction effects to track the evolution of the orbital parameters through time.
Time-domain waveforms produced with this formalism (henceforth referred to as EFB-T) are given by \cite{Loutrel_2020} 

\begin{equation}\label{eq2: EFB-T}
     h_{+, \times}(t) = - \frac{M^2\eta}{p[\ell(t)]d_L} \sum_{k=0}^6 \sum_{n=0}^2 \epsilon^n \Phi^{(n, k)}(\iota, \psi) \, + \mathcal{O}(\epsilon^3)
\end{equation}

where $M$ is the total mass of the binary, $\eta = m_1m_2/M^2$ the \text{symmetric mass ratio}, $\iota$ - $\psi$ the inclination and polarization angles respectively, $p$ is the \text{semi-latus rectum} of the orbit which corresponds to the distance perpendicular to the semi-major axis to one of the focuses. In $G=c=1$ units, it can be measured in $\si{M}_\odot$ units, and it is convenient to normalize it with respect to the total mass $M$: $\bar{p} \equiv p/M$. It is also related to the pericenter distance by 

    \begin{equation}\label{eq2: pericenter distance from semi latus rectum}
        \bar{r}_p = \frac{\bar{p}}{1+e}
    \end{equation}
    
$\ell(t)$ is the \text{mean anomaly} defined as
    \begin{equation}
    \ell(t) = \frac{2\pi}{T_{orb}(t)}(t-t_p)
\end{equation}with $t_p$ the time of periastron passage and $T_{orb}$ the orbital period.
The relation $p[\ell(t)]$ accounts for radiation reaction effects at $2.5$ Post Newtonian order (see Sec. IIIB in \cite{Loutrel_2020}). The waveforms so computed are valid only near $t_p$ ($t\in [-t_{l=\pi}, t_{l=\pi}]$) and reproduce the parabolic limit as $e \rightarrow 1$.

\begin{figure}[!t]
%\hspace{-10mm}
    \subfloat[$10\,\si{M}_\odot + 10\,\si{M}_\odot$]{
        \includegraphics[width = 0.5 \columnwidth]{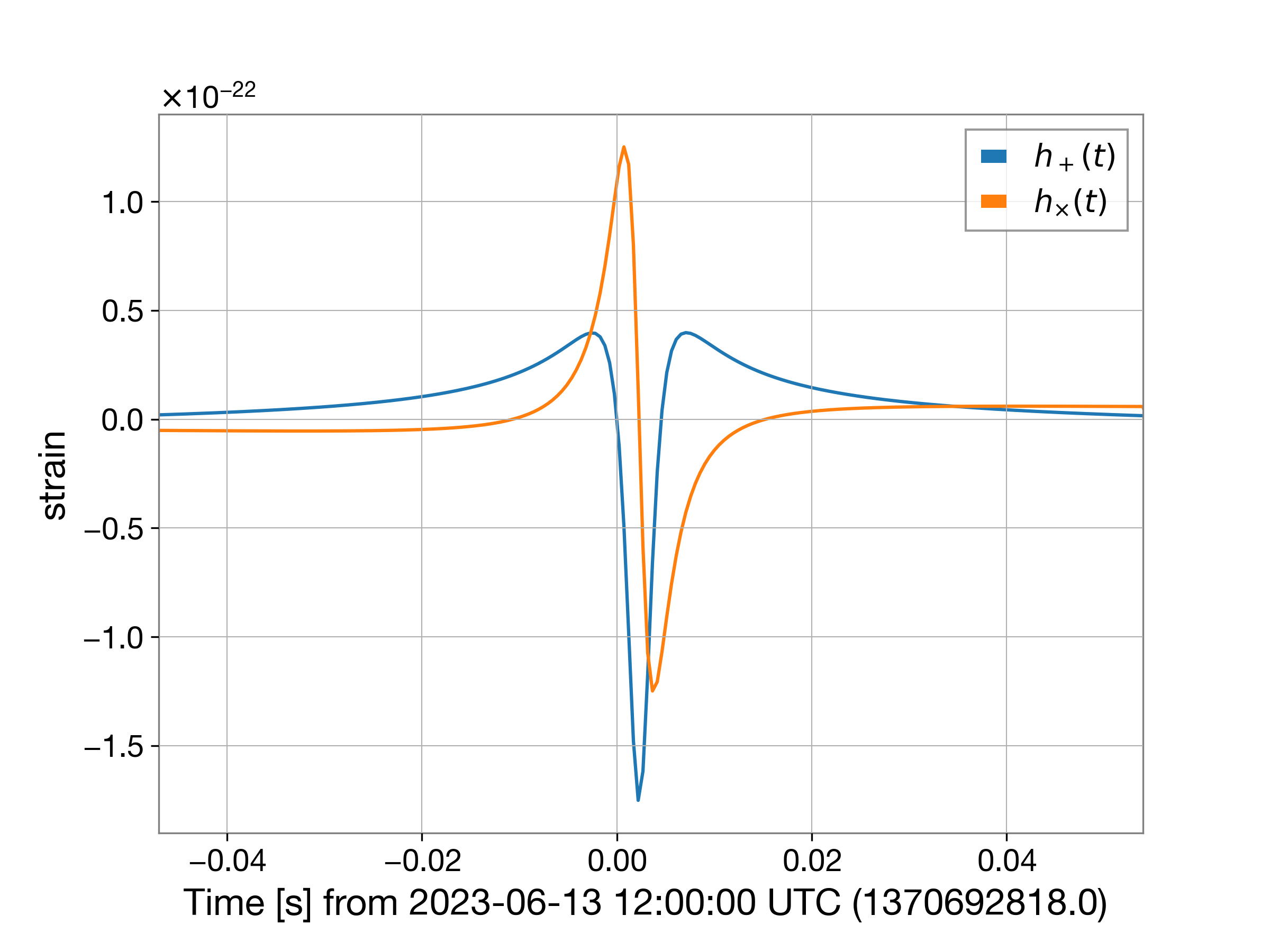}}
    \subfloat[$100\,\si{M}_\odot + 100\,\si{M}_\odot$]{
        \includegraphics[width = 0.5\columnwidth]{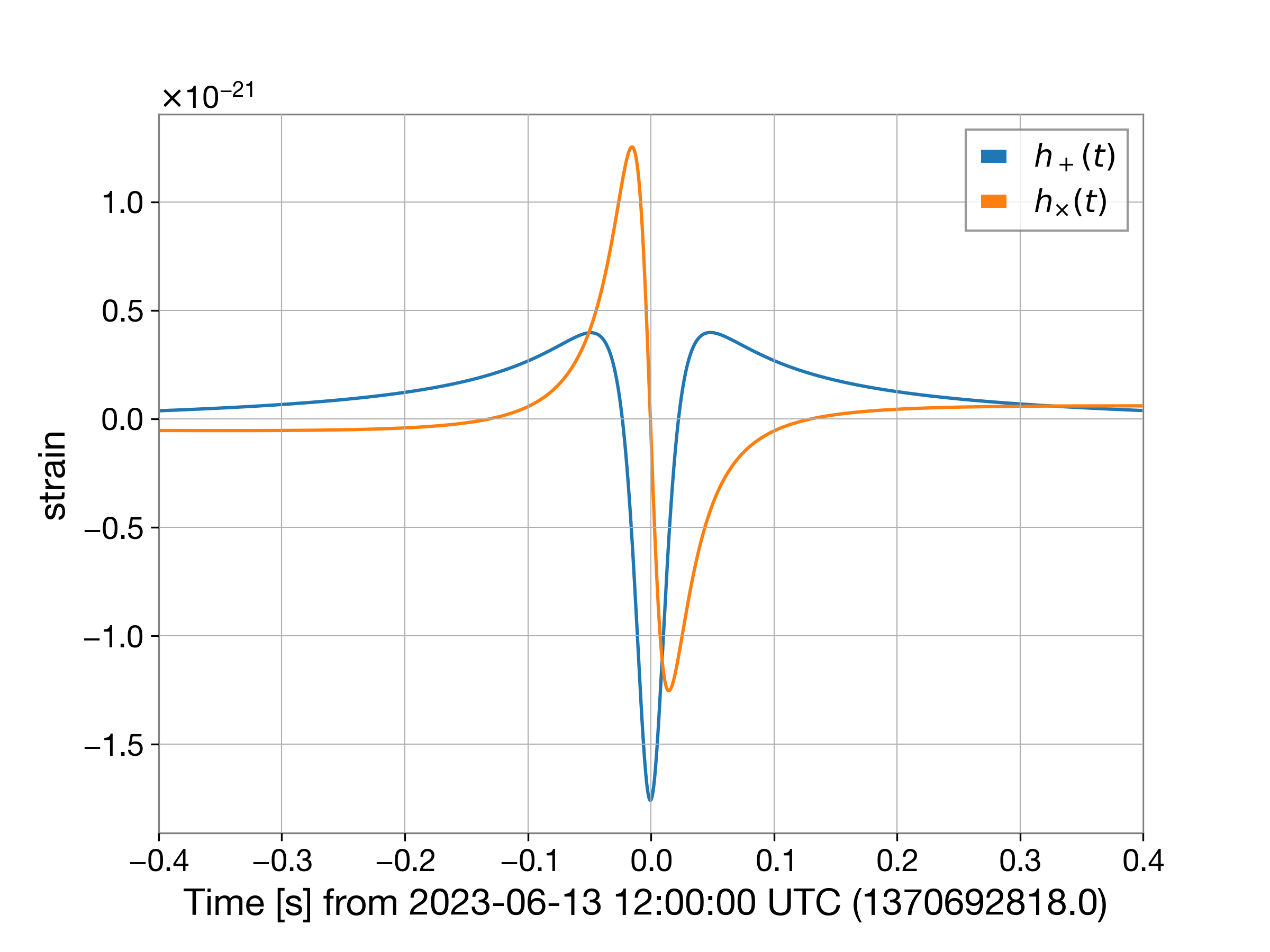}}
        \hfill
    \subfloat[$10\,\si{M}_\odot+ 10\,\si{M}_\odot$]{
        \includegraphics[width = 0.5\columnwidth]{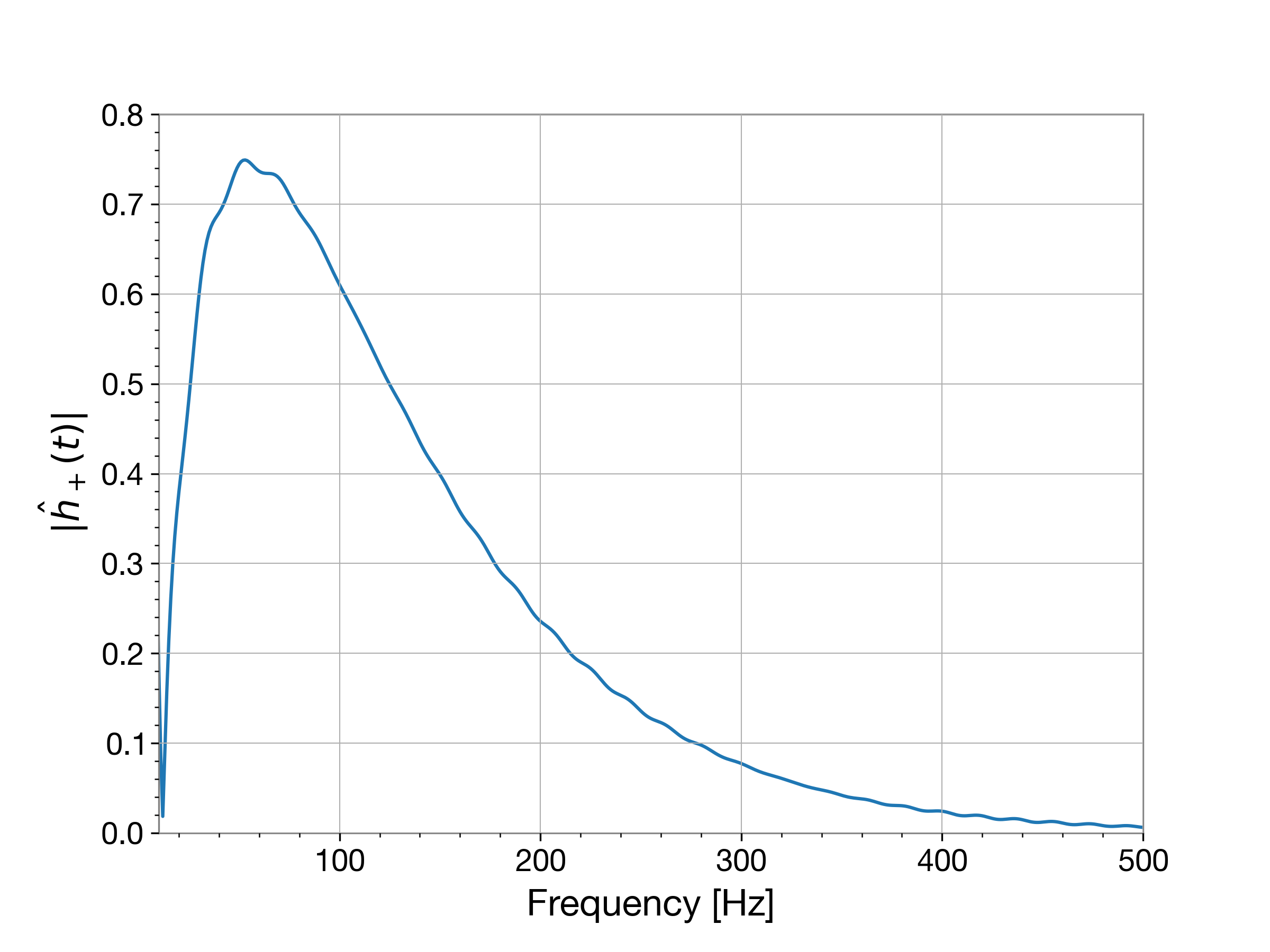}}
    \subfloat[$100\,\si{M}_\odot + 100\,\si{M}_\odot$]{
        \includegraphics[width = 0.5\columnwidth]{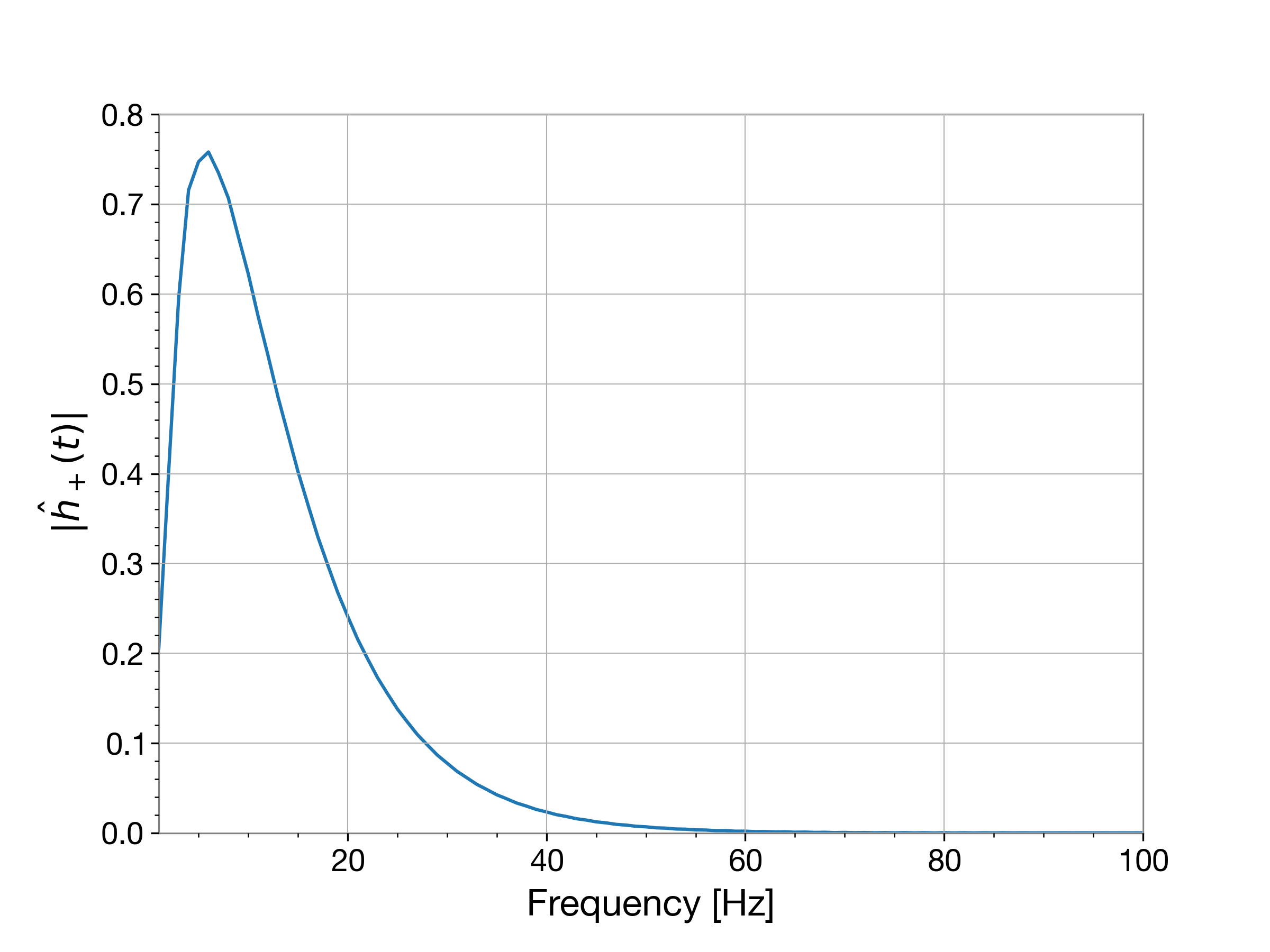}}
        %\caption{}
    \caption{\text{(Top):} plus and cross polarizations waveforms obtained with the Effective Fly-by formalism. The captions indicate the BH masses, while other relevant parameters are $e=0.9$ and $\bar{p}=15$. Note the different timescales. \text{(Bottom):} FFT of the plus polarizations above. We see that the signals lie in the \text{LIGO \& Virgo} sensitivity band and that the increase of the total mass $M$ results in a peak at lower frequencies.}
    \label{fig2: plus&cross polarization + FFTs for EFB-T}
\end{figure}

Examples of the EFB-T plus and cross polarizations waveform are given in Fig. \ref{fig2: plus&cross polarization + FFTs for EFB-T}. From Eq.\eqref[]{eq2: EFB-T} the parameter that mainly affects both polarizations is the total mass $M$. With other parameters fixed, more massive binaries result in a longer and broader burst signal peaked at lower frequencies. Even so, the bursts have very short duration $\lesssim 1$ s, and an overall peak frequency in the range $10-100$ Hz.\\
The good accuracy of these waveforms has been studied in \cite{Loutrel_2020} by comparing it with numerical waveforms at leading Post Newtonian order \cite{Peters&Mathews} and full numerical relativity. 

\section{Normalizing flows for parameter estimation}\label{sec 3: Normalizing Flows}

\subsection{Basic definitions}
The objective of Bayesian Inference in the context of Gravitational Wave data analysis is to obtain the posterior distribution for the parameters describing the signal.  To compute it in the case of Close Encounters sources, we have exploited, in this work, the method of \textit{Normalizing Flows} \cite{papamakarios2021normalizing_flows_for_probabilistic_modeling_inference}.
They are a powerful class of generative models capable of modeling complex probability distributions $p(\vect{\mathrm{x}})$ out of simpler base distributions by means of a learned invertible transformation. The transformation can be conditioned on data thus making it possible to model surrogate posteriors $q(\vect{\theta}|\vect{s}) \approx p(\vect{\theta}|\vect{s})$. The key aspect of this approach is that it does not require any likelihood evaluation as the flow learns how to map $\vect{\theta}$ to the base distribution via a simulation-based process. 
Furthermore, inference requires only to evaluate the inverse transformation on samples from the base distribution, thus leading to a significant reduction in computational inference time.

To introduce the definition of a Normalizing Flow, let $\vect{\mathrm{x}}$ be a vector in an input data space $\mathcal{X}$, distributed as $\vect{\mathrm{x}} \sim p(\vect{\mathrm{x}})$: a  {Normalizing Flow} is then defined by an {invertible map} ({bijection}) $f_\phi: \mathcal{X} \longrightarrow \mathcal{U}$ from the input data space $\mathcal{X}$ to a latent space $\mathcal{U}$ of a random variable $\vect{\mathrm{u}}\sim \pi_{\psi}(\vect{\mathrm{u}})$
\begin{equation}\label{eq4: forward pass}
   \vect{\mathrm{x}} \xrightarrow{\quad f_\phi \quad} \vect{\mathrm{u}} \sim \pi_{\psi} (\vect{\mathrm{u}})\, \quad   \text{(forward pass)}
\end{equation}
Our notation follows \cite{dihn_NICE}, with $\phi$ and $\psi$ parameters $f$ and $\pi$ depend respectively upon. 
Since Eq.\eqref[]{eq4: forward pass} is nothing but a change of variable, the probability distribution $p(\vect{\mathrm{x}})$ can be expressed in terms of the base distribution as:
\begin{subequations}
    \begin{align}
        & p(\vect{\mathrm{x}}) = \pi_{\psi}(\vect{\mathrm{u}}) \left \lvert \det \mathcal{J}_{f_\phi} \right \rvert = \pi_{\psi}(f_\phi(\vect{\mathrm{x}})) \left \lvert \det \left( \dfrac{\partial f_\phi (\vect{\mathrm{x}})}{\partial \vect{\mathrm{x}}} \right) \right \rvert \label{eq4: NF p}\\
        & \log p(\vect{\mathrm{x}}) = \log  \pi_{\psi}(f_\phi(\vect{\mathrm{x}})) + \log  \left \lvert \det \left( \dfrac{\partial f_\phi (\vect{\mathrm{x}})}{\partial \vect{\mathrm{x}}} \right) \right \rvert \label{eq4: NF logp}
    \end{align}
\end{subequations}
where $ \mathcal{J}_{f_\phi} = \left( \dfrac{\partial f_\phi}{\partial    \vect{\mathrm{x}}} \right)$ is the Jacobian of the transformation. \\

The map $f_\phi$ is learned by performing the {forward} pass specified by Eq.\eqref[]{eq4: forward pass}, then the \emph{sampling} of $p(\vect{\mathrm{x}})$ is straightforward and simply consists in evaluating the \emph{inverse} $f_\phi^{-1}$ over samples from the base distribution

\begin{equation}\label{eq4: inverse pass}
   \vect{\mathrm{x}} \xleftarrow{\quad f^{-1}_\phi \quad} \vect{\mathrm{u}} \sim \pi_{\psi} (\vect{\mathrm{u}})\, \quad   \text{(inverse pass)}
\end{equation}
This evaluation can be done as long as some conditions hold. First,  $\pi_{\psi}(\vect{\mathrm{u}})$ must be easy to sample and evaluate. To this scope, the uniform or Gaussian distribution are best suited. Second, $f_\phi$ must be   {invertible}, and third: $f_\phi$ and its inverse are {differentiable}. Furthermore, data and latent spaces share the same   {topology} and   {dimensionality}: the common choice is $\mathcal{X} = \mathcal{U} = \mathbb{R}^D$ 

\subsection{Expressive power and flexibility}\label{subsec4: Flow's Expressive Power and Flexibility }
\begin{figure}[!t]
     \centering
     \includegraphics[width = \textwidth]{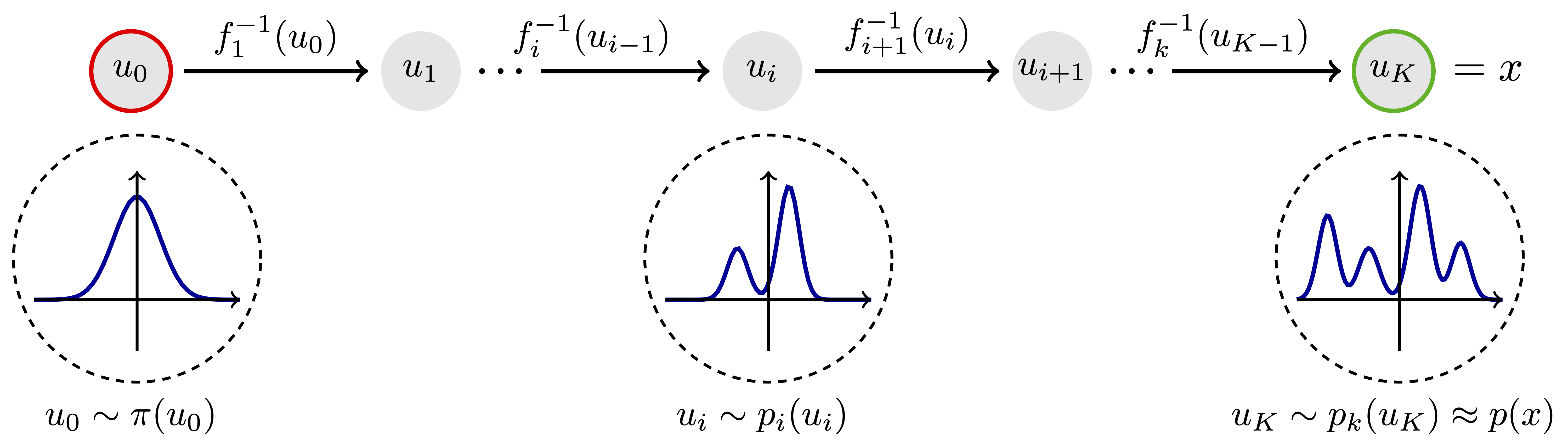}
     \caption{A schematic representation of the   {inverse pass} of a Normalizing Flow where the bijection is made up by a series of composite functions. During the inverse pass (sampling) the samples from the base distribution are gradually transformed in each step into a more complex distribution to match the target. Adapted from \cite{lil_weng2018flow_picture}.}
     \label{fig4: normalizing flow scheme}
 \end{figure}

It is interesting to consider whether a flow-based model can represent any distribution. If $p(\vect{\mathrm{x}})$ and $\pi_\psi(\vect{\mathrm{u}})$ are well behaved distributions satisfying the autoregressivity assumption:
\begin{equation}\label{eq4: autoregressivity assumption}
    p(\vect{\mathrm{x}}) = \prod_{i=1}^D p(x_i| \vect{\mathrm{x}}_{<i})\,, \quad p(x_i| \vect{\mathrm{x}}_{<i})>0 \; \forall i, \vect{\mathrm{x}} \in \mathbb{R}^D
\end{equation}
then  {there exists a diffeomorphism $F$ that can map $\pi_\psi(\vect{\mathrm{x}})$ into $p(\vect{\mathrm{x}})$} \cite{papamakarios2021normalizing_flows_for_probabilistic_modeling_inference}. Although this guarantees its existence, it does not provide a closed formula for $F$, so that it must be learned by optimizing a function $f_\phi$. Therefore the expressive power of a Normalizing Flow, i.e. its ability to model complex distributions, strictly depends on the form of $f_\phi$. 
Making flows more expressive can be achieved by increasing the flexibility of the bijection $f_\phi$. For instance, given that a single function may not be sufficient, the whole bijection can be constructed as a combination of intermediate bijections:

\begin{equation}\label{eq4: multiple layers flow bijection}
    f_\phi = f^{(1)}_{\phi_1} \circ f^{(2)}_{\phi_2} \circ \cdots \circ f^{(K)}_{\phi_K}
\end{equation}
each one with its own set of parameters $\phi_i$ to be optimized. \\
Under this assumption, the Jacobian can be factorized: 
\begin{equation}\label{eq4: jacobian multiple layers}
    \mathcal{J}_{f_\phi} = \prod_{j=1}^K \mathcal{J}_{f^{(j)}_{\phi_j}}
\end{equation}
Then Eq.\eqref[]{eq4: NF logp} reads
\begin{equation}
    \log p(\vect{\mathrm{x}}) = \log  \pi_{\psi}(f_\phi(\vect{\mathrm{x}})) +   \sum_{j=1}^{K} \log  \left \lvert \det \mathcal{J}_{f^{(j)}_{\phi_j}}(\vect{\mathrm{u}}_{j-1}) \right \rvert 
\end{equation}
This shows also the meaning of the name ``Normalizing Flows": the input samples $\vect{\mathrm{x}}$ undergoes a series of composite bijections to be gradually transformed into noise: i.e. $p(\vect{\mathrm{x}})$   {flows} through each discrete step to be  {normalized}. The reverse is true when computing the inverse to sample $p(\vect{\mathrm{x}})$. Fig. \ref{fig4: normalizing flow scheme} gives a graphical representation of this concept.\\
As will be discussed in Sec. \ref{sec4: Constructing Flows}, the bijections may be parametrized with the support of Deep Neural Networks to increase expressiveness.

\subsection{Likelihood-Free Inference}\label{subsec4: likelihood free inference}
The main application of a Normalizing Flow model is   {probability density estimation} and   {sampling}, as stated by Eq.\eqref[]{eq4: NF p}. This approach is useful in cases where it is possible to have access to a collection of samples drawn from an unknown distribution that we would like to reconstruct. Indeed, by fitting the model through Eq.\eqref[]{eq4: forward pass} then new samples can be generated as illustrated by Eq.\eqref[]{eq4: inverse pass}. However the list of possible applications for such models does not end up here, as they can also perform   {variational inference}.
% Cos'è la variational inference?
We will focus more on this kind of application as it fits our studying purposes. \smallskip\\
Our goal is to infer probability distributions for a set of implicit parameters $\vect\theta$ that better describe some observation data $\vect{\mathrm{x}}$. In our particular case $\vect{\mathrm{x}} \equiv \vect{\mathrm{s}}(t)$ is the strain time series containing the gravitational wave signal of a Close Encounter, and $\vect\theta$ the parameters of the physical system that generated it. From the Bayes Theorem

\begin{equation}
    p(\vect{\theta} | \vect{\mathrm{s}}) \propto p(\vect\theta)\, p(\vect{\mathrm{s}}|\vect\theta)
\end{equation}

The   {posterior} distribution $p(\vect{\theta} | \vect{\mathrm{s}})$ is traditionally computed either with Monte Carlo Markov Chain (MCMC) 
or Nested Sampling by repeated evaluations of the   {likelihood} $p(\vect{\mathrm{s}}|\vect\theta)$. This can become a bottleneck in many situations, either because the likelihood function can be costly to evaluate or because it may be not well defined thus preventing a tractable computation.
Alternatively, Normalizing Flows provide themselves as a natural method to approximate the posterior by producing a  {surrogate posterior} $q(\vect{\theta} | \vect{\mathrm{s}})$ in a tractable way. This can be done by making the bijection   {conditioned} on the observed data.

\begin{equation}\label{eq4: NF likelihood free inference}
      {
    p(\vect{\theta} | \vect{\mathrm{s}}) \approx q(\vect{\theta} | \vect{\mathrm{s}}) = \pi_{\psi}(f_\phi(\vect\theta, \vect{\mathrm{s}})) \left \lvert \det \left( \dfrac{\partial f_\phi (\vect\theta, \vect{\mathrm{s}})}{\partial    \vect{\mathrm{\theta}}} \right) \right \rvert 
    }
\end{equation}
It is worth emphasizing that Eq.\eqref[]{eq4: NF likelihood free inference} does not require any likelihood evaluation to perform inference. The only requirement is to be able to   {simulate} the data from a given set of parameters $\vect\theta^*$ extracted from a prior distribution:

\begin{equation}\label{eq4: simulations}
    \vect\theta^* \sim p(\vect{\theta}) \quad,\quad \vect{\mathrm{s}}^* \sim p(\vect{\mathrm{s}}|\vect\theta^*)
\end{equation}

By this simulation process, the model indirectly incorporates both the prior over the parameters and the likelihood since the data points are generated accordingly. Therefore, this whole approach goes under the name of   {Likelihood-Free Inference} or even   {Simulation-based Inference} \cite{simulation_based_inference}. As in other methods based on Machine Learning, inference is significantly faster since the computational cost is mostly during the training phase.

\subsection{Training of Normalizing Flows}\label{sec4: loss function}
The training of Normalizing Flow models consists in optimizing the set of parameters $\vect\phi$ upon which the bijection $f_\phi$ depends by minimizing a suitable loss function. 
Given our purposes of inferring a surrogate gravitational wave posterior, in order for $q(\vect{\theta} | \vect{\mathrm{s}}) \approx  p(\vect{\theta} | \vect{\mathrm{s}})$ it is necessary to minimize the distance between the two. The most straightforward measure of how close two distributions are is the {Kullback-Leibler divergence} \cite{KL_divergence}. The true posterior is in principle unknown but we can use the simulated set of samples $\{\vect\theta^{(i)}, \vect{\mathrm{s}}^{(i)} \}_{i=1}^N$ to minimize the \emph{forward} KL divergence $\mathbb{KL}[p||q]$. \\
\indent It is possible to derive an expression for the loss in the following way
\begin{equation}\label{eq4: NF loss}
    \begin{split}
        \mathcal{L} &= \mathbb{KL}\left[\,p(\vect{\theta} | \vect{\mathrm{s}})\,||\,q_{\vect\phi}(\vect{\theta} | \vect{\mathrm{s}})\,\right] = \\[10pt]
        & = \int d\vect{\mathrm{s}}\, p(\vect{\mathrm{s}}) \int d\vect\theta \,p(\vect{\theta} | \vect{\mathrm{s}}) \log \left( \frac{p(\vect{\theta} | \vect{\mathrm{s}})}{q_{\vect\phi}(\vect{\theta} | \vect{\mathrm{s}})} \right) = \\
        & = \int d\vect{\mathrm{s}}\, p(\vect{\mathrm{s}}) \left[ - \underbrace{\int d\vect\theta\, p(\vect{\theta} | \vect{\mathrm{s}})\: \log q_{\vect\phi}(\vect{\theta} | \vect{\mathrm{s}})}_{\mathbb{H}[p||q_\phi]}   +  \underbrace{\int d\vect\theta\, p(\vect{\theta} | \vect{\mathrm{s}})\: \log p(\vect{\theta} | \vect{\mathrm{s}})}_{\mathbb{H}[p||p] = cost}    \right] \simeq \\[20pt]
        & \simeq - \int d\vect\theta \: p(\vect\theta) \int d\vect{\mathrm{s}} \:p(\vect{\mathrm{s}}|\vect\theta )\: \log q_{\vect\phi}(\vect{\theta} | \vect{\mathrm{s}}) \simeq\\[10pt]
        &\simeq - \frac{1}{N} \sum_{i=1}^N \log q_{\vect\phi}(\vect{\theta}^{(i)} | \vect{\mathrm{s}}^{(i)})
    \end{split}
\end{equation}
where $\mathbb{H}[p||q_\phi]$ is the differential Cross-Entropy between the two distributions and $\mathbb{H}[p||p]$ can be discarded, being constant with respect to the flow's parameters. At the fourth line we have applied the Bayes Theorem to express the cross entropy in terms of the likelihood instead of the unknown posterior and in the last passage we leveraged the fact that we are in a simulation based context (cf. Eq.\eqref[]{eq4: simulations}) which implies that the integral can be approximated via Monte Carlo methods.

Therefore, minimizing the KL divergence is equivalent to minimizing the Cross-Entropy between $p$ and $q$.
By substituting Eq.\eqref[]{eq4: NF logp} we obtain the final formula

\begin{equation}\label{eq4: NF complete loss}
      { \mathcal{L} = - \frac{1}{N} \sum_{i=1}^N \left[\;\log  \pi_{\psi}(f_\phi(\vect{\mathrm{\theta}}^{(i)}; \vect{\mathrm{s}}^{(i)})) + \log  \left \lvert \det \left( \dfrac{\partial f_\phi (\vect\theta^{(i)} ;\vect{\mathrm{s}}^{(i)})}{\partial \vect{\mathrm{\vect \theta}}} \right) \right \rvert \;\right]
    }
\end{equation}

So, minimizing the loss defined by Eq.\eqref[]{eq4: NF complete loss} guarantees that the distribution inferred by the flow will converge to an optimal approximation for the true posterior. That relation remarks another time the   {likelihood-free} nature of this approach since even to optimize the flow no likelihood evaluations are required: hence, the likelihood enters in the model via the simulated dataset. \\
Furthermore, the optimization of the parameters can be performed via Stochastic Gradient methods since unbiased estimators for the gradients are given by \cite{papamakarios2021normalizing_flows_for_probabilistic_modeling_inference}:

\begin{subequations}
    \begin{align}
        & \nabla_\phi \mathcal{L} \approx  - \frac{1}{N}\sum_{i=1}^N \nabla_\phi \pi_{\psi}(f_\phi(\vect{\mathrm{\theta}}^{(i)}; \vect{\mathrm{s}}^{(i)})) + \nabla_\phi \log \lvert \mathcal{J}_{f_\phi}(\vect\theta^{(i)} ;\vect{\mathrm{s}}^{(i)}) \rvert \label{subeq4: gradient for f_phi}\\
        & \nabla_\psi \mathcal{L} \approx  - \frac{1}{N}\sum_{i=1}^N \nabla_\psi \pi_{\psi}(f_\phi(\vect{\mathrm{\theta}}^{(i)}; \vect{\mathrm{s}}^{(i)}))\label{subeq4: gradient for base distribution}
    \end{align}
\end{subequations}
Eq.\eqref[]{subeq4: gradient for base distribution} is due to the fact that in some applications the base distribution can be learned together with the flow as well. However in the case of likelihood-free inference is common practice to keep it fixed.
In deriving Eq.\eqref[]{eq4: NF complete loss} we opted to minimize the {forward} KL divergence $\mathbb{KL}\left[\,p\,||\,q\,\right]$.
In principle, there are other possible divergence measures that can be minimized: here, we motivate our choice. An alternative could have been the {reverse} KL divergence $\mathbb{KL}\left[\,q\, ||\,p\,\right]$. This is typically adopted when the target density $p$ is easy to evaluate but difficult to sample, which is not our case with posteriors over gravitational wave parameters. There is, however, a much more profound reason why the reverse is not the best option. First of all, KL divergence is \emph{not} symmetrical. Thus, minimizing either one or the other leads to different results, as the optimized distribution will show different behaviours. More specifically, the   {forward} KL is   \emph{mass covering} while the reverse is   \emph{mode seeking}. 
An intuitive explanation can be suggested. In the forward case, in order for KL to not diverge, $q>0$ whenever $p>0$, meaning it must cover the whole support of $p$. Conversely, in the reverse case, being $p$ at the denominator: $q=0$ whenever $p=0$, thus forcing $q$ to seek for the dominant mode in $p$.
In the case of a multimodal distribution, as gravitational wave posteriors are, a mass covering approximant is preferable since it will not exclude less dominant modes that could provide interesting information.

\subsection{Normalizing Flows for gravitational wave data analysis}\label{sec4: NFs for gw data analysis}
Two main algorithms are currently exploited to infer the Bayesian posteriors over gravitational wave parameters: MCMC and Nested Sampling. Both are based on Markov Chains and obtain samples from $p(\vect\theta | \vect{\mathrm{s}})$ by means of repeated likelihood evaluations. This implies several computational drawbacks.
First of all, the computational efficiency of these algorithms is severely limited by waveform generation, which can take about $10^{-3}{s} \lesssim \braket{\tau} \lesssim 1{s}$ \cite{parallel_bilby}, depending on the particular waveform model used. 
This, combined with the elevated number of required likelihood evaluations $\mathcal{O}(10^7)$ \cite{Number_of_likelihood_evaluations_gw}, gives a hint about the amount of time required to perform an analysis. Secondly, being based on Markov Chains,  the produced samples show correlation, which has to be accounted for, thus reducing the number of {effective samples}. The high inference time is perhaps the most relevant limitation since it also impacts Multimessenger observations as an {Early Warning} strategy is hardly implementable.
Furthermore, the typically adopted Gaussian likelihood (see, e.g., \cite{LALInference}) assumes Gaussian (Wide sense) Stationary Noise in the detector. Such a condition is not always completely satisfied as detectors may manifest both non-gaussianities and non-stationarities like the frequent short transients known as {glitches}. Therefore, if the noise assumptions are violated, the whole analysis can be affected by biases.

Parameter estimation analyses typically require a precise knowledge of the waveform models. In the case of Close Encounters, where uncertainties exist on the waveform modeling, it has been shown that the recovery of parameters (e.g., the masses) is limited by a small number of accessible bursts during the inspiral \cite{detecting_and_PE_CE}. A NF-based approach can leverage the generalization capabilities of deep neural networks to better recover the parameters with a limited amount of information, providing, at the same time, reduced inference times.\\
Finally, computational efficiency will become a key aspect of data analysis in future observing runs as well as in the third-generation detector era. As a consequence of the higher sensitivity of future instruments, it is expected a $\sim$10$^{3}$ increase in the event rate $\mathcal{R}$. As an example, $\mathcal{R} \gtrsim 10^5$ events/year for the Einstein Telescope \cite{ET_science_case}. Faster and more efficient algorithms will be crucial for the success of those experiments.\\
Normalizing Flows provide themselves as a valid alternative able to supply to the limitations of traditional methods. In fact, as we discussed in Sec. \ref{subsec4: likelihood free inference}, the cost of inference is completely amortized as likelihood evaluations are not required, and expensive waveform computations are performed only once during training. The fact of being a simulation-based inference has another implication worth emphasizing: it does not suffer from the limiting assumption about gaussianity and stationarity of the noise,
provided that an adequate description is available.

\subsection{Model Selection with Normalizing Flows}\label{subsec4: model selection with NFs Importance Sampling}
Another kind of analysis that strictly depends on Parameter Estimation is   {model selection} (or   {hypothesis testing}), which in the case of gravitational waves may refer to signal detection, i.e. testing the hypothesis of the presence or absence of a signal in the strain, or even discriminating between two waveform models what is better at describing the data. 
This is done by computing the Bayes factor $\mathcal{B}_{12} = {\mathcal{Z}_1}/{\mathcal{Z}_2}$ which compares the evidences (or marginal likelihoods) of the two hypothesis. Furthermore, when computed in the case of the null hypothesis of having only noise ($\mathcal{Z}_2 = \mathcal{L}_{noise}$), $\mathcal{B}_{12}$ can be exploited as a detection statistic.\\
Although the product of a Normalizing Flow model is a direct approximation of the posterior, the evidence can be estimated as well through   {Importance Sampling}, which is nothing but a Monte Carlo estimate. More precisely
\begin{equation}
    \mathcal{Z} = \int d\vect\theta \:p(\vect\theta)\, p(\vect{\mathrm{s}}|\vect \theta) = \int d\vect\theta \:\frac{p(\vect\theta)\, p(\vect{\mathrm{s}}|\vect \theta)}{q(\vect \theta | \vect{\mathrm{s}})}\,q(\vect \theta | \vect{\mathrm{s}})
\end{equation}
By sampling the flow posterior $q(\vect \theta | \vect{\mathrm{s}})$, which is optimized by minimizing the mass covering forward KL divergence, we can get an estimator of the evidence from importance sampling weights.
\begin{equation}\label{eq3: model evidence NF Importance Sampling}
    \hat{\mathcal{Z}} = \frac{1}{N} \sum_{i=1}^N \frac{p(\vect\theta_i)\, p(\vect{\mathrm{s}}|\vect \theta_i)}{q(\vect \theta_i | \vect{\mathrm{s}})} = \frac{1}{N} \sum_{i=1}^N w_i
\end{equation}
The only disadvantage is that $w_i$ relies on the analytical likelihood to be computed. However, since they can be computed separately, the whole procedure can be parallelized in principle, reducing its computational cost.

\subsection{Constructing the Flow}\label{sec4: Constructing Flows}

We now discuss how Normalizing Flows can be constructed by implementing the bijection $f_\phi$ to be   {expressive} and   {computationally efficient} at the same time.
When referring to computational efficiency, the interest is to find a function whose Jacobian, actually its determinant, is fast to compute. The function must also be easy to invert and rapid to evaluate both in the forward and inverse pass. On the other hand   {expressiveness} refers to a sufficiently flexible transformation able to deal with highly complex distributions. More in general since $f_\phi: \mathbb{R}^D \rightarrow \mathbb{R}^D$ acts on $D$-dimensional vectors, it has the general form:
\begin{equation}\label{eq4: bijection general form}
    u_i= g_{\phi_i}(\theta_i; \vect \Theta_i)\;, \quad \vect \Theta_i = c_i (\vect{\mathrm{\theta}})
\end{equation}
where $c_i$ is the \textit{conditioner}, which specifies how the bijection acts on the various dimensions and, in particular, on which set of components $\vect \Theta_i$ does $\theta_i$ depends. It is not required for it to be a bijection.  $g_\phi$ is instead the   \textit{transformer}: a monotonic function, hence invertible, that actually transforms the input variables.
The set of parameters $\vect\phi$ of $f_\phi$ contains both parameters of the conditioner and transformer. Since, however, the conditioner is typically specified before the training and it is not part of the optimization, its are just {hyperparameters}. For this reason, henceforth, we'll refer to $\vect\phi$ as the parameters of the transformer only. 

The most simple flow that can be constructed is the so-called {\textit{Element-wise Flow}} whose conditioner treats each vector dimensions independently Eq.\eqref[]{eq4: element-wise flow}. 
\begin{equation}\label{eq4: element-wise flow}
    \begin{cases}
         u_1 = g_{\phi_1}(\theta_1)  \\
         u_2 = g_{\phi_2}(\theta_2)  \\
         \quad\;\;\: \vdots \\
         u_D = g_{\phi_D}(\theta_D)
          
    \end{cases}\;, \quad \det \mathcal{J}_{f_{\phi}} = \prod_{i=1}^D \dfrac{\partial g_{\phi_i}}{\partial u_i}
    %\mathcal{J}_{f_{\phi}}(\vect\theta) = \left[\begin{array}{cccc}
    %   \dfrac{\partial g_{\phi_1}}{\partial \theta_1}   &  0 & \cdots& 0 \\
    %    0   &  \dfrac{\partial g_{\phi_2}}{\partial \theta_2} &   & \vdots \\
    %    \vdots   &   & \ddots  & 0 \\
    %    0   & \cdots  & 0  & \dfrac{\partial g_{\phi_D}}{\partial \theta_D} \\
    %\end{array}\right]
\end{equation}

This flow is efficient both in the forward and inverse pass due to the simple Jacobian: being a diagonal matrix, its determinant is just the product of the diagonal. However, it lacks expressiveness since each component is transformed independently. Hence, it won't be able to capture all the eventual dependencies and degeneracies among the various elements. In the case of gravitational waves, there are a lot of dependencies between parameters. As an example, recall from Eq.\eqref[]{eq2: EFB-T} that in the case of Close Encounters, the strain amplitude is $h_{+, \times}(t) \propto M^2/d_L$ which induces a degeneracy between the total mass and the luminosity distance. Other degeneracies can arise, for instance, when considering the localization of the source and the antenna pattern response of the detectors. There are other architectures able to deal with such situations, like the   \emph{Autoregressive} conditioner or the  \emph{Coupling Layers}. 

The former, in particular, models the dependencies between variables assuming an autoregressive structure  Eq.\eqref[]{eq4: autoregressivity assumption} where each component $\theta_i$ depends upon $\theta_{j<i}$ components. With this assumption the bijection Eq.\eqref[]{eq4: bijection general form} becomes

\begin{equation}\label{eq4: autoregressive flow transform}
    u_i = g_{\phi_i}(\theta_i)\quad \text{with} \quad \phi_i = F(\vect{\theta}_{1:i-1})
\end{equation}

In most cases, $g_\phi$ is taken to be an analytical invertible function whose parameters $\phi$ are the output of a Neural Network here denoted with $F$ \cite{MAF}. 
The autoregressive transformation Eq.\eqref[]{eq4: autoregressive flow transform} is characterized by having a low triangular Jacobian 

\begin{equation}
    \mathcal{J}_{f_\phi}(\vect\theta) = \left[\begin{array}{ccc}
      \dfrac{\partial \mathrm{u}_1}{\partial \theta_1}   &  & \vect0\\
         & \ddots &\\
       \vect{\mathrm{A}}  &  &\dfrac{\partial \mathrm{u}_D}{\partial \theta_D} 
    \end{array}\right] 
\end{equation}
hence making the computation of its determinant equivalent to the Element-wise Flow Eq.\eqref[]{eq4: element-wise flow}.

Nevertheless, the whole architecture of Autoregressive Flows manifests inefficiency when computing the inverse transformation (inference) as it takes a \emph{recursive} structure. In fact, the sampling of $\theta_i$ from $\vect{\mathrm{u}}$ requires to have already sampled $\vect\theta_{1:i-1}$ thus turning this operation in a sequential and {non} parallelizable one: see e.g. Fig. 3 in \cite{papamakarios2021normalizing_flows_for_probabilistic_modeling_inference}. 
The computational cost scales in particular as $\mathcal{O}(D)$. 
It is, therefore, an unavoidable aspect of Autoregressive Flow to have either one of the two passes to be inefficient\footnote{It has been proposed indeed a slight variation of this flow which is the {Inverse Autoregressive Flow} \cite{IAF}. The recursive structure is moved from the inverse to the forward pass, but it can't be removed.}. Although they are, in principle, the most expressive since they are able to account for any dependence in the variables, the computational cost either for training or sampling scales badly with high dimensional inputs.

{Coupling Layers} were introduced in \cite{dihn_NICE} to overcome the efficiency limitations of Autoregressive Flows while maintaining their expressiveness.
The idea behind a coupling layer is to split the parameter space in two equally dimensional subsets $\vect\theta = (\vect\theta^d, \vect\theta^{D-d})$ with $d\simeq D/2$. The second half is then transformed element-wise and conditioned on the first half, which is mapped through an identity.

\begin{equation}\label{eq4: coupling layer}
    \begin{cases}
        \vect{\mathrm{u}}_{1:d} = \vect\theta_{1:d}\\
         \vect{\mathrm{u}}_{d+1:D} = g_\phi(\theta_{d+1:D}; \vect \theta_{1:d})
    \end{cases}
\end{equation}

The Jacobian is still a low triangular matrix
\begin{equation}
    \mathcal{J}_{f_\phi}(\vect\theta) = \left[\begin{array}{cc}
      \mathbb{I}_d    & \vect0\\
         &\\
       \vect{\mathrm{A}}    &\vect{\mathrm{D}}_{D-d}
    \end{array}\right] \;, \quad \vect{\mathrm{D}} = \diag \left[ \frac{\partial \mathrm{u}_{d+1:D}}{\partial \theta_{d+1:D}} \right]
\end{equation}

However, since the upper left block is simply the identity matrix, the computational cost scales as $\mathcal{O}(D-d)$. It turns out that computing both forward and inverse is an efficient operation that can be further parallelized. The only drawback of this layer is that a single one is not sufficient, as only half of the components get actually transformed. To enhance the expressiveness, it is possible to stack multiple of these layers with random permutations of $\vect\theta$ indexes in between. Hence, $f_\phi$ is given by Eq.\eqref[]{eq4: multiple layers flow bijection} and the full Jacobian by Eq.\eqref[]{eq4: jacobian multiple layers}. If the number $K$ of layers is sufficient, the output will be equivalent to an autoregressive one due to the fact that, in the end, each component is transformed, being conditioned on every other component. As a ``rule of thumb", $K$ should at least be equal to $D$. \\
\indent Coupling Layers provide themselves as the optimal choice both in terms of expressiveness, flexibility, and computational cost: in fact, both training and sampling are equally fast. Moreover, they are also relatively easy to implement. 

We now describe the invertible transformation. Any strictly monotonic function, being invertible, can be applied, provided, however, it is differentiable and with an easy-to-compute inverse. 
In the continuation of this discussion, we will consider two of the most widely adopted.
\emph{Affine transformations} were among the first functions to be proposed as suitable transformers. The same work introducing coupling layers  adopted this form exploiting exponential rescaling \cite{dihn_NICE,RealNVP}:

\begin{equation}\label{eq4: RealNVP - Affine Transformation}
      {
    \begin{cases}
    \vect{\mathrm{u}} = g_\phi(\vect \theta) \;\:= \vect\theta \odot \exp \left[s(\vect \theta)\right] + t(\vect \theta) \\
    \vect\theta = g^{-1}_\phi(\vect{\mathrm{u}}) = \left[ \vect{\mathrm{u}} - t(\vect{\mathrm{u}}) \right] \odot \exp \left[-s(\vect{\mathrm{u}})\right] 
    \end{cases}
    }
\end{equation}
In Eq.\eqref[]{eq4: RealNVP - Affine Transformation}, $\odot$ denotes the Hadamard (or element-wise) product.
The parameters of this function are hence $\phi = \{t, s\}$: shift and scale.\\
Combined with Coupling Layers, the transformation of Eq.\eqref[]{eq4: RealNVP - Affine Transformation} 
has proven to be flexible and expressive enough to model complex distributions as images \cite{GLOW} or even audio waveforms \cite{kim2019flowavenet}. 

Another flexible transformation was introduced in \cite{neural_spline_flow} as an avenue to model extremely complex and multimodal distributions while retaining the property of being analytical, differentiable, and easy to invert. 
The idea is to map an interval $[-B, B]\subset \mathcal{X}$ into $[-B, B] \subset \mathcal{U}$ by interpolating a {Rational Quadratic Spline} between a set of sorted knots $\{x_k, y_k\}_{k=0}^K$ where both the knots and their internal derivatives $\{\delta_k\}_{k=1}^{K-1}$ are parametrized as the output of a Neural Network. Computing the inverse requires solving a $2^{{nd}}$ order equation, which can be done analytically (see Eqs. (6)$-$(8) in \cite{neural_spline_flow}).
This kind of transformation is extremely flexible, and it naturally induces multimodality by increasing the number $K$ of knots. Therefore, it has been mainly applied in the context of image generation. 

\section{HYPERION's Architecture}\label{Description of Architecture}
\begin{figure}[!ht]
    \centering
    \includegraphics[width = 0.7\textwidth]{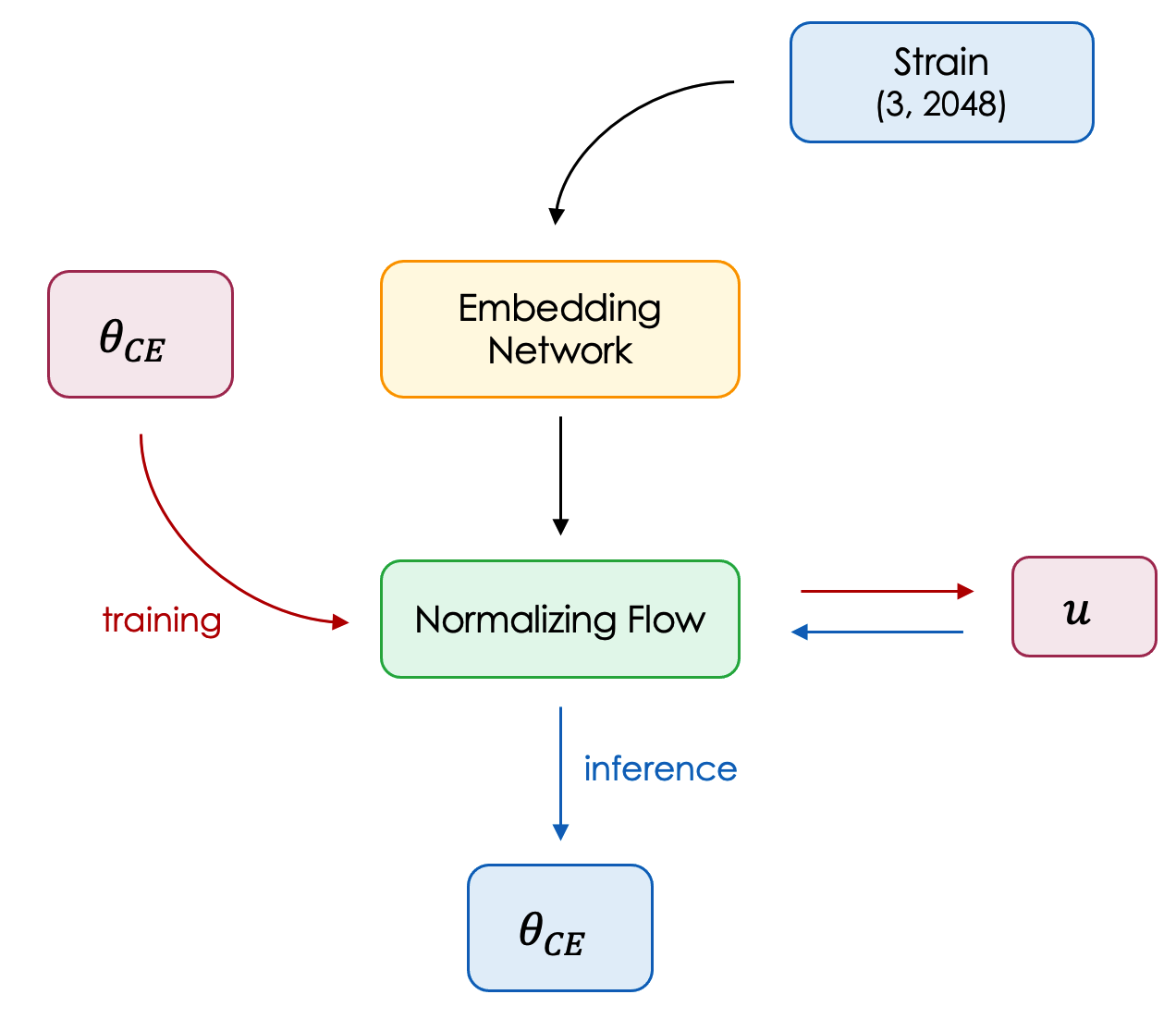}
        \caption{Schematic overview of \textsc{HYPERION} which is composed of a \text{Normalizing Flow} and and \text{Embedding Neural Network} acting on input strain data. Solid arrows represents \text{input-ouput} relations: \text{red} apply during training, \text{blue} ones when performing inference while \text{black} ones are always present. \label{fig5: HYPERION Model structure}}
\end{figure}

We present here the ‘‘\textbf{HYP}\textit{er-fast close }\textbf{E}\textit{ncounte}\textbf{R I}\textit{nference from} \textbf{O}\textit{bservations with }\textbf{N}\textit{ormalizing-flows}" pipeline (HYPERION). This pipeline takes as input \SI{1}{\second} of whitened strain time series and returns as output samples from the posterior probability $p(\vect{\theta}_{CE}|\vect{s})$.
More specifically, the parameters over which it makes inference are: the total mass $M$, mass ratio $q$, eccentricity $e_0$\footnote{the subscript 0 refers to the value when the mean anomaly $\ell = 0$, i.e., the periastron passage}, semi-latus rectum $\bar{p}_0$, luminosity distance $d_L$, the time of periastron passage $\delta t_p$, right ascension $\alpha$ and declination $\delta$. The general structure of \textsc{HYPERION} is depicted in Fig. \ref{fig5: HYPERION Model structure} along with \text{input/output} relations between its building blocks.

The core of the model is a \text{Normalizing Flow}, which reconstructs the posterior distribution.
Given that it is a conditional probability distribution, the flow must be supplied with the most informative context as possible. Therefore, we introduced in the model another building block, fundamental as well: an \text{Embedding Neural Network}. Acting as a \text{feature extractor}, its primary task is to extrapolate the information in the noisy strain time series and to compress it to a lower dimensional form. This procedure has the purpose of \text{filtering out} all the irrelevant features, mainly the noise content. 
Other than the Embedding one, other Deep Neural Networks are implemented in the Normalizing Flow itself, thus making our model reach the number of $\sim 180$ millions of trainable parameters. \textsc{HYPERION} was developed with \texttt{python} 3.10 and \textsc{PyTorch} 2.1.0 \cite{pytorch}.

\subsection{The Embedding Network}\label{subsec5: embedding network}

\begin{figure}[!ht]
    \subfloat[\label{fig5: embedding network}]{
  \includegraphics[width=\columnwidth]{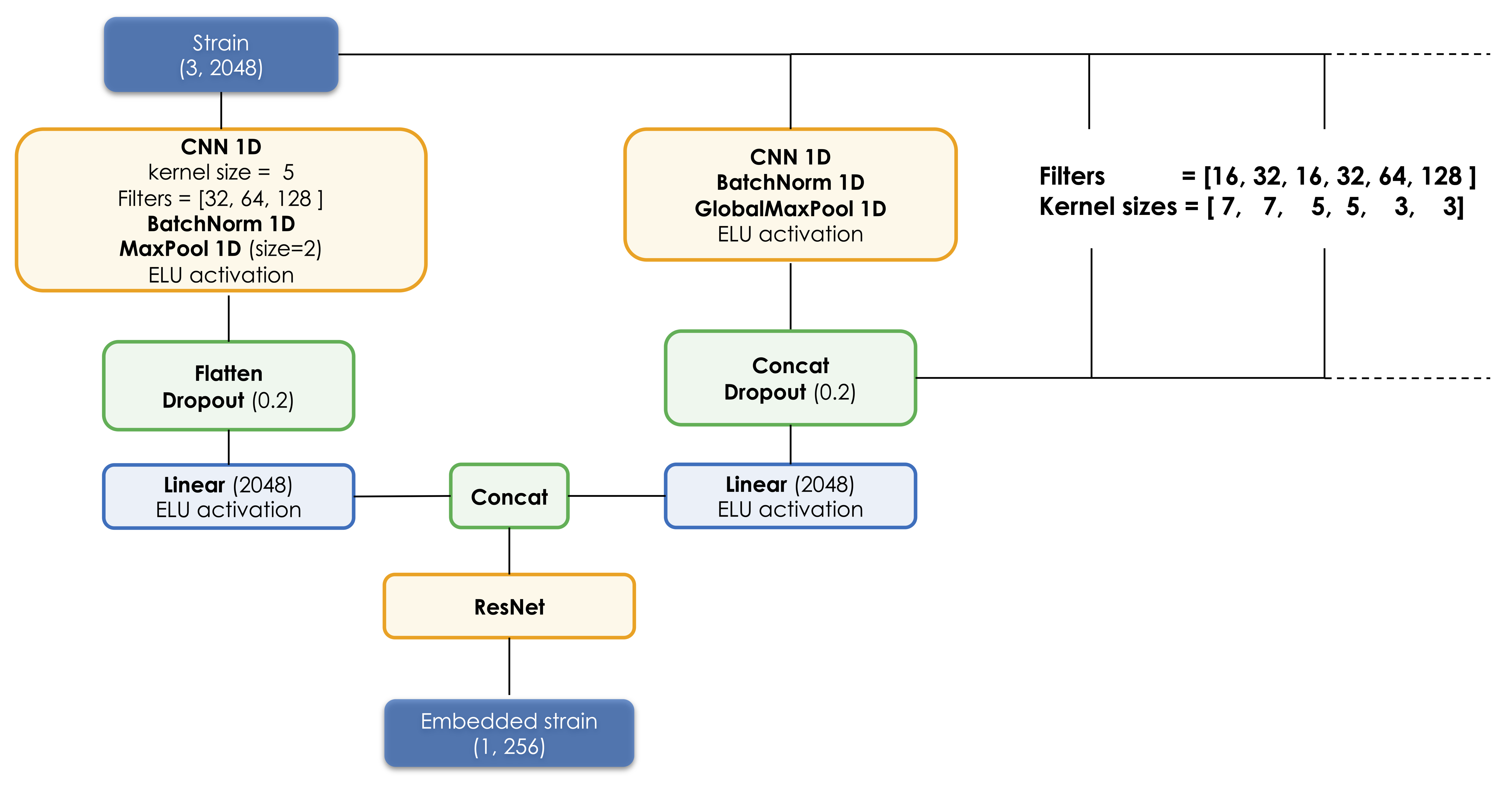} }\hfill
    \subfloat[\label{fig5: resnet block}]{
  \includegraphics[width=\columnwidth]{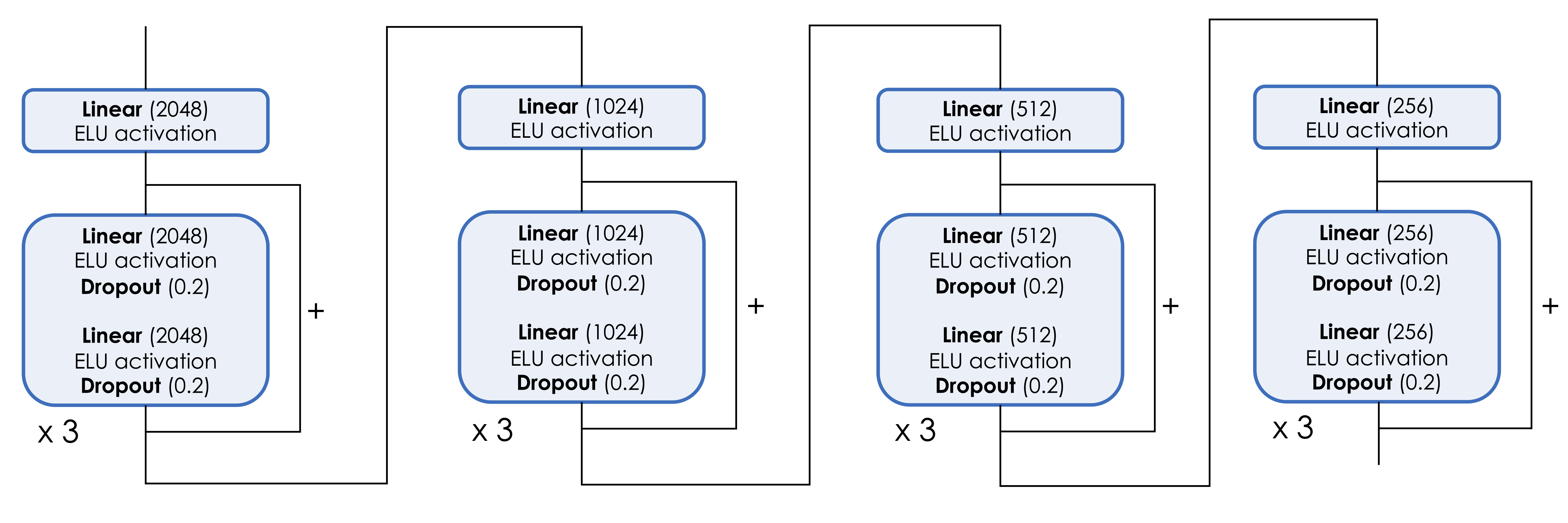} }\hfill
    
    \caption{\text{(a):} The general architecture of the \text{Embedding Network} is composed of two CNN blocks acting in different ways and a ResNet block that efficiently compresses the extracted features into a $(1, 256)$ dimensional tensor. More specifically, the CNN block on the left extracts features related to the signal morphology, while the other on the right focuses more on temporal correlated patterns. \text{(b):} Detailed architecture of the ResNet block. \label{fig5: embedding + resnet block}} 

\end{figure}

The presence of such an element in our model can be justified by the following reason. In the process of \text{Likelihood-free Inference}, the likelihood enters indirectly as the result of a simulation procedure. It means that the NF is able to determine the best mapping $f_\phi:\Theta \rightarrow \mathcal{U}$ based on the \text{similarity} between the joint samples $\{\vect{\theta}^{(i)}_{CE}, \vect{\mathrm{s}}^{(i)}\}$ that it is supplied with. A raw data representation, like the strain time series, is not the optimal choice, even if whitened. That is both because of the low signal to noise ratio for CE signals and because of the morphology of the signal itself, which does not show directly a clear dependency on all the $\vect{\theta}_{CE}$ parameters. Hence, a feature extractor is necessary. 
The overall architecture of the Embedding Network shown in Fig. \ref{fig5: embedding network} is the result of several optimizations and improvements.  

It is composed of two Convolutional Neural Network (CNN) blocks that perform the \text{feature extraction} from the input time series: $1 \text{s}$ sampled at $f_s=2048\text{ Hz}$ with each of the 3 channels corresponding to a given interferometer (H1, L1, V1). The first block consists of three 1D Convolutional Layers with fixed \text{kernel size} $=5$ and $[32, 64, 128]$ numbers of filters respectively. Between each layer, there are pooling layers and also \text{Batch Normalization} layers, whose addition was found to be beneficial. This first block is able to learn features related to the shape or morphology of the signal which are relevant for a subset of $\vect{\theta}_{CE}$. Given, however, the translational invariance of the neuron's response in such block, it is unable to learn features that are mostly correlated with their. 

Indeed, in our earlier experiments, the inference about $\{\delta t_p, \alpha, \delta\}$ was not great as we were simply recovering their priors. The reason for the effect on sky localization can be easily understood by the fact that it is determined by the relative shifts in arrival time of the signal in each detector
. For long signals, like those produced by coalescences, multiple time instants can be compared: i.e., the information spreads over a wide temporal interval. For a burst signal, on the contrary, all the emission is concentrated over a small time interval, which implies that the sky localization information strongly correlates with $\delta t_p$ itself.

This motivated introducing a parallel CNN block acting differently from the former. The crucial difference is that many convolutional layers, with different filters and kernel sizes, slide independently over the time series and after each of them a \text{Global Max Pooling} layer keeps the maximum neuron's response. The resulting outputs are then concatenated together, and in this way, the temporal information about the neuron's response gets preserved.  

The output of each CNN block is then passed to Linear layers with 2048 neurons that are subsequently concatenated and then compressed into a final layer with $256$ output neurons by means of the ResNet block (Fig. \ref{fig5: resnet block}). This block is composed of four sub-blocks sized [2048, 1024, 512, 256] respectively, each one containing 3 skip connections. In contrast to regular linear layers, skip connections proved to be more efficient at compressing the dimensionality of the network's output without loss of meaningful information.
Regarding activation functions, we have found the best results with the ELU rather than ReLU.  To reduce at minimum the chances of overfitting the Embedding Network, especially the ResNet block, makes extensive use of Dropout layers

We decided to exploit CNNs and not Recurrent Neural Networks (RNNs) mainly for two reasons. RNNs are suited for the analysis of long temporal correlated sequences. In our case, as already explained, the information is localized in time. CNNs are better suited to extract feature on different timescales, therefore they have been proposed as a viable machine learning method for gravitational wave data analysis. Furthermore, the recurrent structure of RNNs negatively impacts the computational cost of inference.

\subsection{The Normalizing Flow}\label{subsec5: NF architecture}

\begin{figure}[!t]
    \centering
    \includegraphics[width=\textwidth]{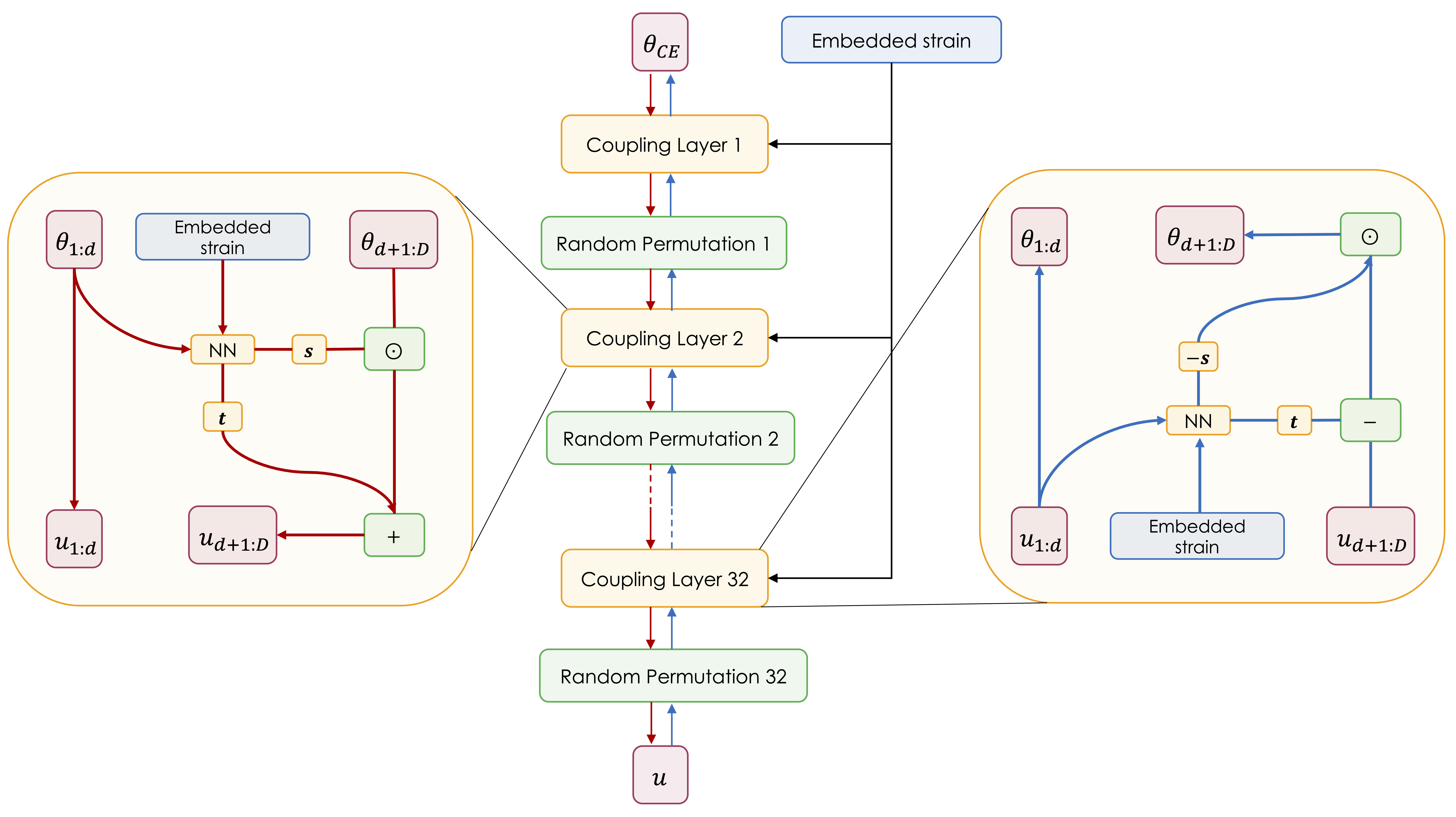}
    \caption{The architecture of the Normalizing Flow implemented in \textsc{HYPERION} which consists of a stacking of 32 coupling layers with affine transformations. The red arrows refer to training, while the blue ones to inference (inverse of the transformation). }
    \label{fig5: HYPERION NF scheme}
\end{figure}

The NF implemented in HYPERION adopts \text{Coupling Layers}, given their properties and computational efficiency, combined with affine transformations Eq.\eqref[]{eq4: RealNVP - Affine Transformation}. The whole Normalizing Flow scheme is shown in Fig. \ref{fig5: HYPERION NF scheme} where we made explicit the structure of the Affine Coupling Layer Eq.\eqref[]{eq5: affine coupling layer}.   

\begin{equation}\label{eq5: affine coupling layer}
    \begin{split}
        (\text{training}):\qquad &  \begin{cases}
            u_{1:d} = \theta_{1:d} \\
            u_{d+1:D} = \theta_{d+1:D} \odot \exp{ \left[ s_{d+1:D}(\theta_{1:d}) \right] } + t_{d+1:D}(\theta_{1:d})
        \end{cases}    \\[0.5ex]
        (\text{inference}):\qquad &  \begin{cases}
            \theta_{1:d} = u_{1:d} \\
            \theta_{d+1:D} = \left[u_{d+1:D} - t_{d+1:D}(u_{1:d)}) \right]\odot \exp{ \left[ -s_{d+1:D}(\theta_{1:d}) \right] } 
        \end{cases}  
    \end{split}
\end{equation}

The architecture consists of  32 layers: a sufficiently high number to guarantee a proper mixing between all the $\vect{\theta}_{CE}$ components in order to capture all the dependencies and degeneracies. In between every coupling layer, a \text{Random Permutation} shuffles the parameter's space indexes. This can be seen as an additional transformation with a Jacobian equal to $\mathbbm{1}$. The permutation matrices are then saved as parameters of the model for reproducibility. We also tested Rational Quadratic Splines, given their expected expressiveness, but they turned out to be sub-optimal.
The posteriors produced were excessively multimodal, with clear signs of either underfitting or overfitting in some cases: this was also confirmed by the training and validation losses during the optimization.

\begin{figure}[!t]
    \centering
    \includegraphics[width = 0.85\textwidth]{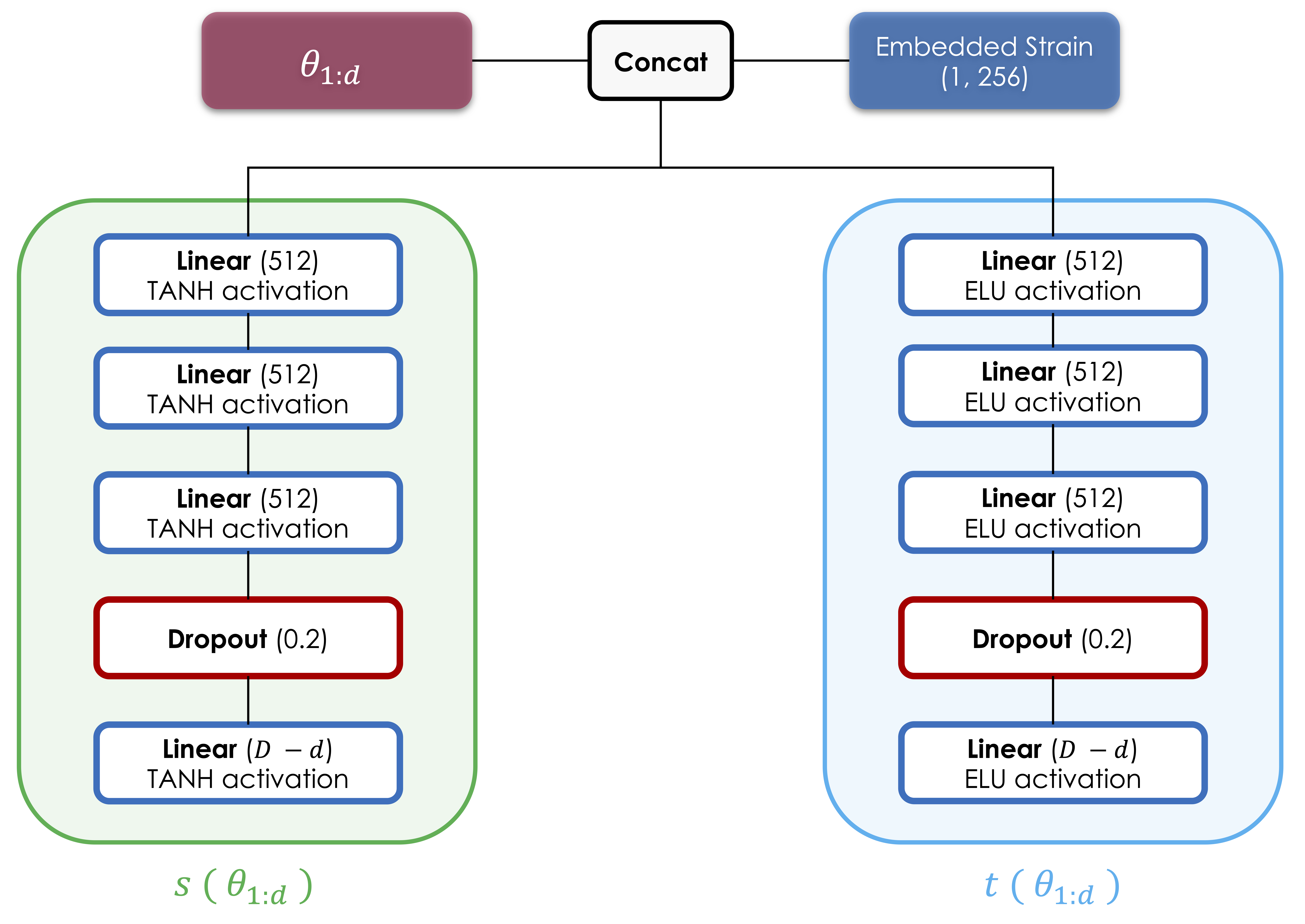}
    \caption{Neural Network architecture for the Affine Coupling Layer Eq.\eqref[]{eq5: affine coupling layer}. Each of the 32 Coupling Layers in \textsc{HYPERION} contains such a Network that gets optimized independently. Note the different activation functions for the scale parameter branch.}
    \label{fig5: flow network}
\end{figure}

The Affine Couplings depend upon two parameters: scale $\vect{s}$ and shift $\vect{t}$. Both of them are the output of a Fully Connected Neural Network (Fig. \ref{fig5: flow network}), which takes as input both the identity-mapped parameters and the embedded strain. In our implementation, each layer has its own Network that is optimized independently for an overall more precise inference. The Network for the scale and shift parameters are nearly identical except for the activation function. While the shift's one adopts the ELU, the other one adopts the $\tanh$ to prevent numerical instabilities that can arise otherwise due to the fact that $\vect{s}$ enters into an exponential. The hyperbolic tangent is also a better choice than the Sigmoid since it allows both $\leq 1$ and $\geq 1$ scale factor values.  

%\newpage
\newpage

\subsection{Pre and Post Samples Processing}\label{subsec5: pre and post processing}

Since the various parameters in $\vect{\theta}_{CE}$ might have wide and different numerical ranges, a direct usage of their strict value would certainly result in numerical instabilities when fed to the Neural Networks. For that reason, each parameter is rescaled to have zero mean and unit variance. This reduces their numerical range while keeping intact the shape of their prior distribution at the same time. Means $\vect{\mu}$ and standard deviations $\vect{\sigma}$ are computed from the training dataset and saved as model hyperparameters. At the end of the inference phase, all the samples are brought back to their original physical range.

\section{Simulations and results} \label{Simulations and results}
\subsection{Training Dataset}
Since \text{Likelihood-free inference} with Normalizing Flows relies on simulated training data samples that must reflect the properties of real ones, the simulation of the training dataset is one of the most delicate operations of this work.
The dataset is made up by joint samples $\mathcal{D}= \{\vect{\theta}_{CE}^{(i)}, \,\vect{\mathrm{s}}^{(i)}\}_{i=1}^{N}$ where $\vect{\theta}_{CE}$ are the Close Encounter gravitational wave signal parameters and $\vect{\mathrm{s}}$ is the corresponding \text{strain time series} sampled at $\SI{2048}{\;Hz}$. 
This sampling frequency implies a \text{Nyquist} frequency $f_{nyq} = \SI{1024}{\;Hz}$ large enough to capture the frequency spectrum of CE BBH, whose peak frequency is in the band $10-100 \text{ Hz}$.
We prepared a dataset of $N = 5 \times 10^6$ samples, being the best compromise between an accurate inference and computational training cost.
The first step in generating the dataset is the sampling of $\vect{\theta}_{CE}$ from prior distributions. Those parameters are then fed into the Effective Fly-by model to produce plus and cross template polarizations $h_{+, \times}(t)$. Depending on the source sky coordinates, the template is afterward projected onto Advanced LIGO and Virgo detectors.\
This simulated signal is embedded into 8 seconds of Gaussian colored noise sampled from the reference O3a amplitude spectral density and saved in a Hierarchical Data Format file. We allowed the amplitude spectral density to vary for each simulated event in order to reproduce the non-stationarity of background noise. 

In this work, we have not included transient noises like \text{glitches} since the capability of analyzing the time series of three detectors simultaneously automatically rejects local sources of noise.
This whole procedure is parallelized, therefore significantly reducing the simulation time to $\mathcal{O}(10 \text{ h})$ on a \textsc{AMD Epyc 7301}CPU with 32 cores / 64 threads.

\begin{figure}[!ht]
    %\hspace{-5mm}
    \subfloat[]{\includegraphics[width = 0.5\textwidth]{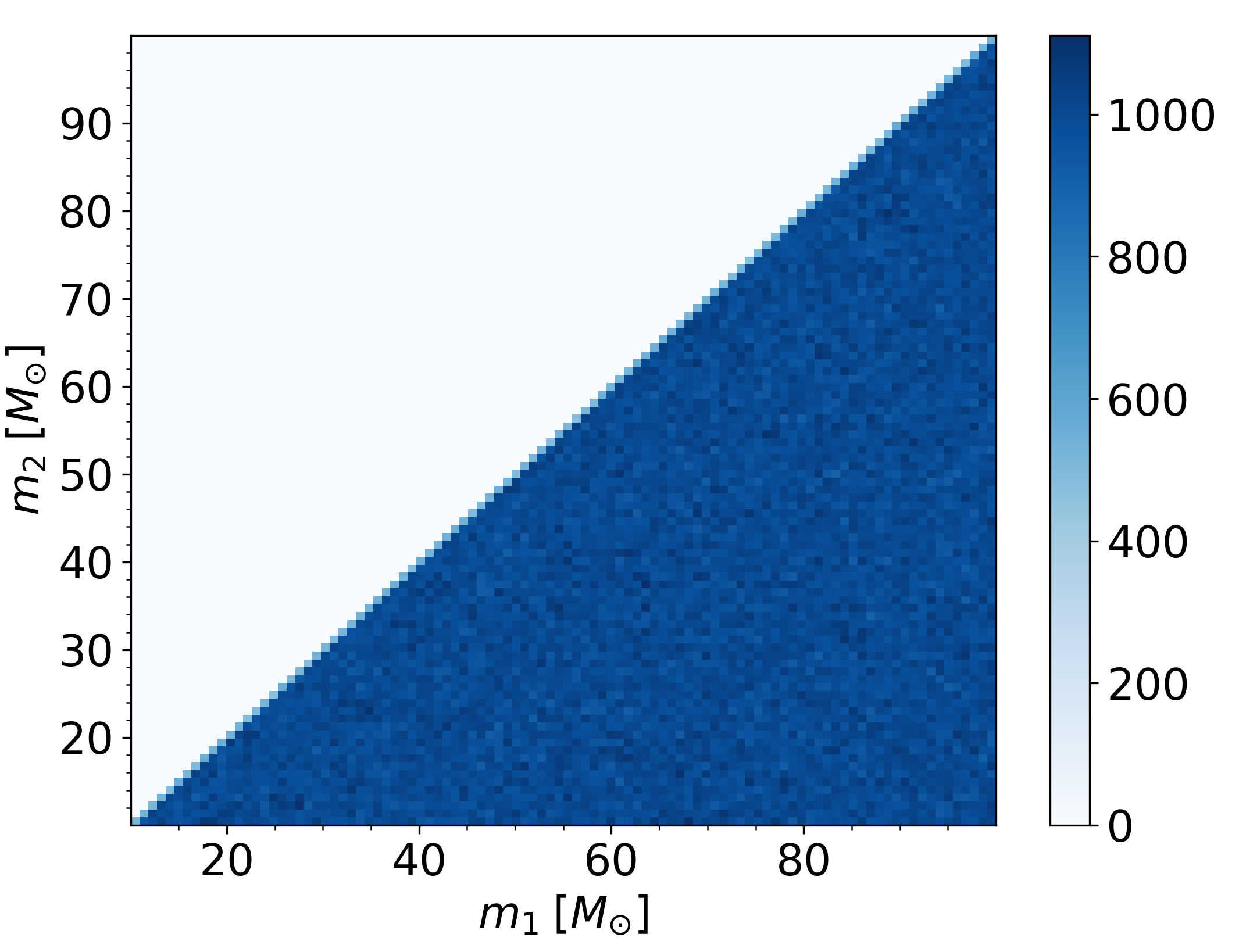}}
    \subfloat[]{\includegraphics[width = 0.5\textwidth]{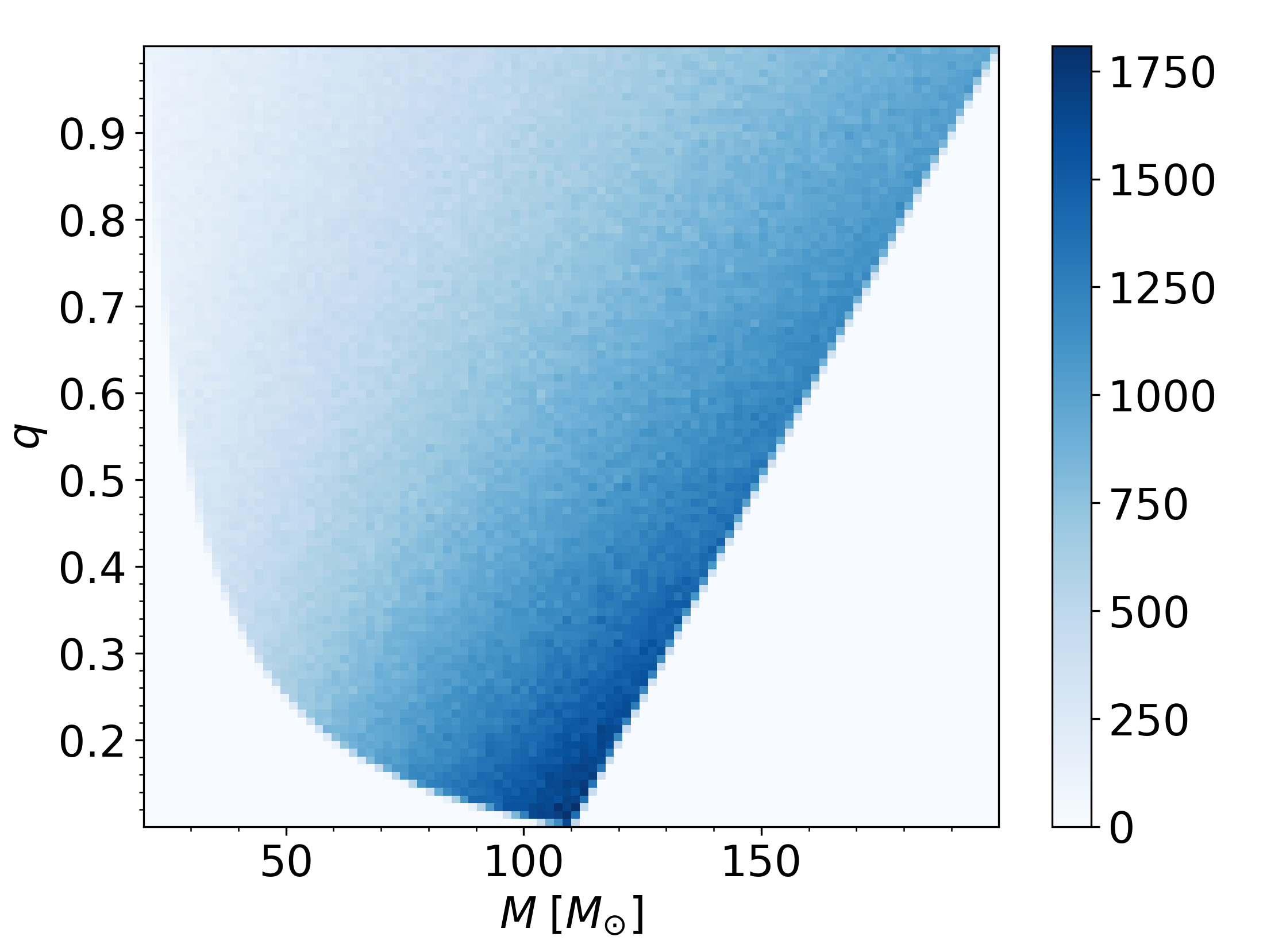}}
    \caption{Prior distributions over the mass parameters. \text{(a):} uniform prior over $m_i$ with the condition $m_2 \leq m_1$. This is the prior implemented in the simulations. \text{(b):} The same mass prior but in terms of \text{total mass} $M = m_1+m_2$ and \text{mass ratio} $q= m_2/m_1$. 
    Instead of the two mass components, \textsc{HYPERION} makes inference on $M$ and $q$.}
    \label{fig5: mass priors}
\end{figure}

The prior distributions over $\vect{\theta}_{CE}$ are listed in Tab. \ref{tab5: CE prior distribution}.
\begin{table}[!ht]
\centering
\begin{tabular}{cccc}
\hline
\hline%\\ %[-1.5ex]
$\theta_{CE}$ & distribution & min        & max    \\ %[1.5ex] 
\hline
\hline %\\
    $m_1$ [$M_\odot$]     &      uniform        &      $10$      &  $100  $         \\
      $m_2 \leq m_1$ [$M_\odot$]     &      uniform        &      $10 $     &  $100    $       \\
     $\bar{p}_0$     &     uniform         &   $ 13 $       &  $  25  $      \\
      $d_L$ [Mpc]    &      uniform         &     $ 100  $    &   $2000  $       \\
      $e_0$     &     uniform      & $  0.85  $       &  $   0.95  $     \\
       $\alpha$   &        uniform       &     $0$       &     $2\pi$       \\
        $ \delta$    &       $\cos$       &      $-\pi/2$      &      $\pi/2$      \\
$\delta t_p$ [s]      &     uniform       &   $-0.25$         &   $0.25$         \\%[1.2ex]
\hline 
%\\[0.8ex]
  $\psi$      &    uniform     &     $0$       &     $\pi$       \\
    $\iota$            &      $\sin$        &     $0$       &   $\pi$         \\
GPS time      &     fixed  & \multicolumn{2}{c}{$1370692818.0$} \\%[1.2ex]
\hline
\hline
\end{tabular}
\caption{Prior distributions of the simulated BBH CE population. The first set of parameters is the one over which \textsc{HYPERION} makes inference, while the rest enters only in the simulation phase. $\alpha$ and $\delta$ are \text{right ascension} and \text{declination} respectively.}
\label{tab5: CE prior distribution}
\end{table}
These population parameters can be grouped into the following categories:
 
\textit{Mass components:} we adopted a uniform prior over $m_{1, 2}$ (Fig. \ref{fig5: mass priors}). As the strain amplitude in the \text{EFB-T} model Eq.\eqref[]{eq2: EFB-T} scales with the total mass $M$, the model makes inference on $M = m_1+m_2$  and $q = m_2/m_1$. The condition $m_2\leq m_1$ reduces the number of effective simulations and hence computational resources; 

\textit{EFB-T parameters:} namely the \text{eccentricity} $e_0$, \text{semi-latus rectum} $\bar{p}_0$ (normalized with the total mass $M$) and time of peri-astron passage $\delta t_p$ with respect to a reference GPS time. The ranges for these parameters are chosen for the template waveforms to provide the highest match with Numerical Relativity. In particular, we adopt the same prior choices of \cite{Gleetter_Nunzio}; 

\textit{Gravitational wave localization parameters:} the sky angles $\alpha$ (RA) and $\delta$  (DEC) whose prior is chosen to be \text{uniform over the sphere}. For the \text{luminosity distance} $d_L$ we chose the range $100$ Mpc $- \,2$ Gpc. We opted for a uniform prior to produce a more balanced dataset. It's worth to note also that, since from Eq.\eqref[]{eq2: EFB-T} $h(t) \propto M^2/d_L$ a biased estimate of $d_L$ could indeed introduce a bias also in the estimate of $M$; 

\textit{Other gravitational wave parameters:} additional parameters relevant for the simulations of the gravitational wave emission are the GPS time at which the event occurs, which is fixed for all the simulations, \text{polarization} angle $\psi$ and \text{inclination} angle $\iota$ between the orbital angular momentum and the line of sight. For the last two, we adopt standard physical priors. At the moment, these parameters are not included in the inference process.

\subsection{Training procedure}
We trained the model for 250 training epochs, each one ending after the flow has been optimized over 1000 batches made of 512 samples.\\
We used the ADAM optimizer \cite{kingma2017adam} with an initial learning rate of $10^{-4}$. During the training, 10\% of the dataset was reserved for validation, and the learning rate was reduced by 50\% after 10 epochs without validation loss improvements. 
\begin{figure}[!t]
    \centering
    \includegraphics[width=\textwidth]{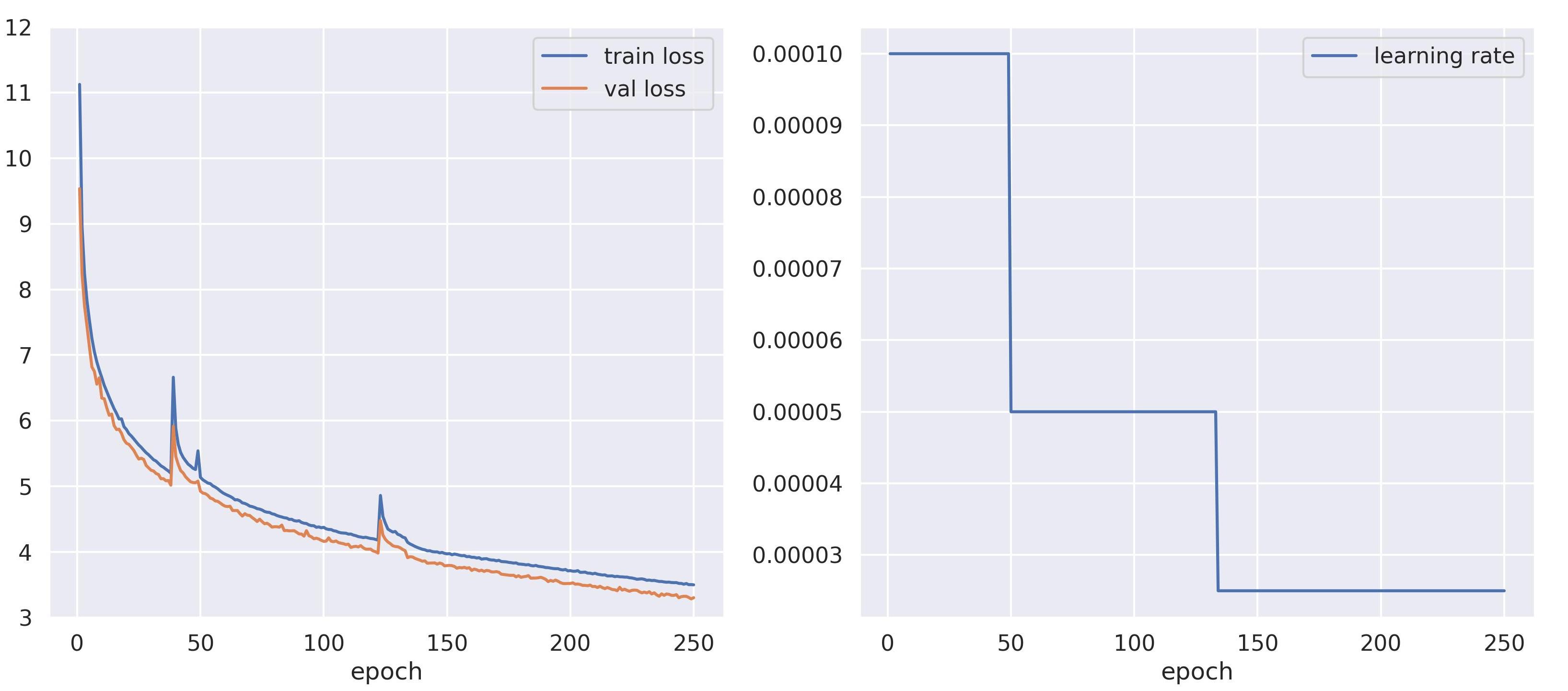}
    \caption{Plot of the training history. \text{(Left):} Training and Validation Loss over the 250 training epochs. The close agreement between the two indicates no issue of overfitting. \text{(Right):} Learning Rate schedule during training.}
    \label{fig5: history plot training validation loss}
\end{figure}
Before training, the training dataset is completely simulated and preprocessed. In particular, during the preprocessing phase, the strain is whitened and cropped to one second. No highpass filter was applied to avoid the risk of cutting out relevant signal frequencies. 
During training, we did not apply any augmentation except for the time of periastron passage $\delta t_p$, which is randomly drawn from the prior (Tab. \ref{tab5: CE prior distribution}) for any training sample loaded into the GPU. The relative strain time series is rolled accordingly. Given the short duration of the signal, there is no risk for it to get too close to the time series edges. This augmentation has proven to be quite effective at reducing overfitting. 
The training history is shown in Fig. \ref{fig5: history plot training validation loss}. As both the training and validation loss are in close agreement, we conclude there is no sign of overfitting.

The value of 512 for the batch size is an optimal compromise between the training stability and final model accuracy.
Moreover,  the usage of only 1000 batches for optimization during each epoch ensures a good covering of the training sample's parameter space without the model having seen the whole dataset. This strategy is similar to the one adopted in \cite{Gabbard_2021_Vitamin_CVAE}.

Tuning the learning rate $\eta$ was a crucial aspect of the training phase. The typical starting value of $\eta_0 = 10^{-3}$ has been demonstrated to be too large and did not allow a proper optimization. We have also tested different annealing strategies, like the \textit{cosine annealing}, although we found best results with the strategy oulined earlier.
\\
The whole training phase took around 20 hours on a Dell PowerEdge R7425 machine equipped with NVIDIA A30 GPUs. 
 
\subsection{Performance on parameter inference}
\begin{figure}[!ht]
    \centering
    \includegraphics[width = 0.9\textwidth]{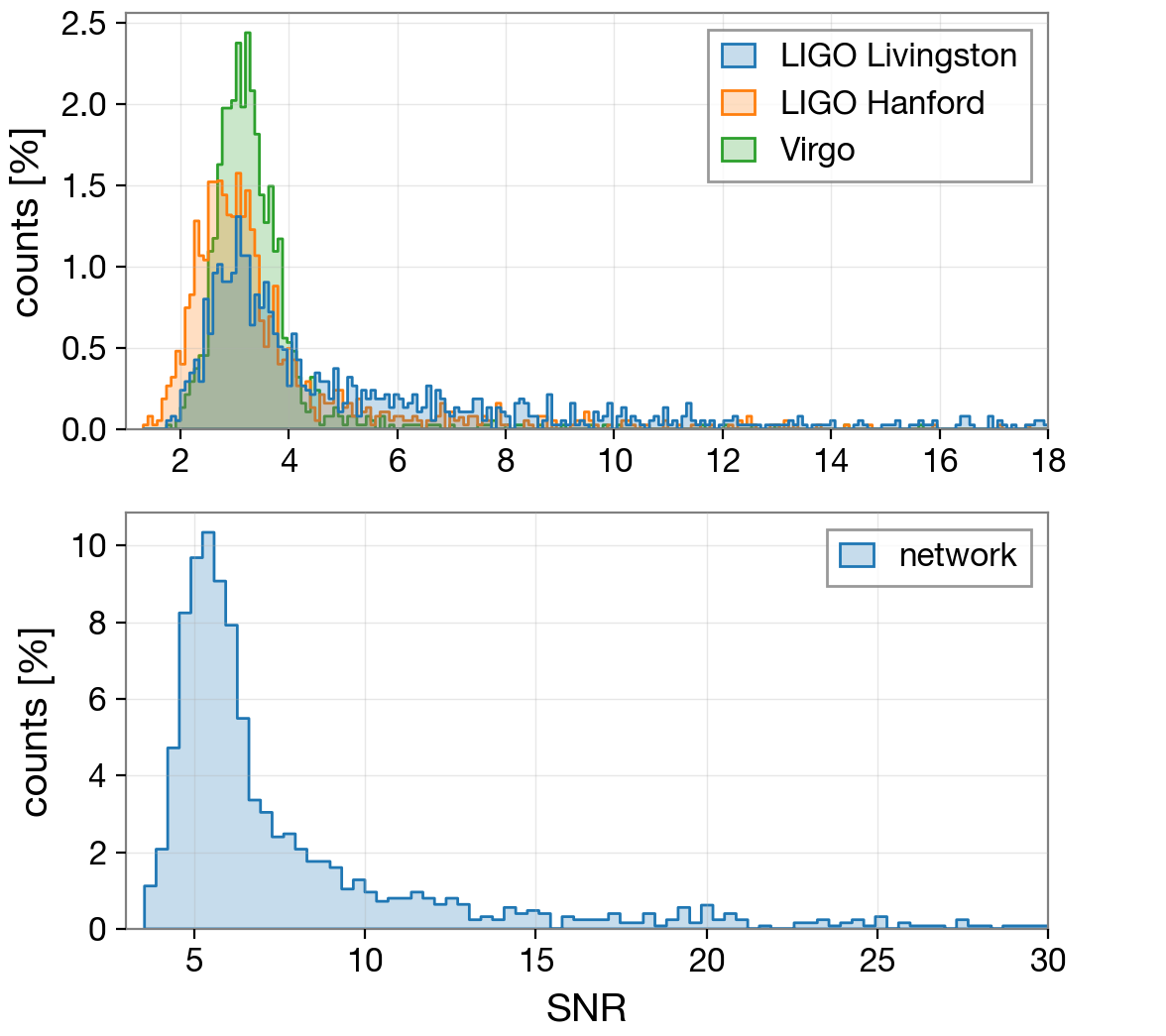}
    \caption{SNR distribution of the signals in the Test Dataset. \text{(Top):} SNR distribution for each of the simulated detectors. \text{(Bottom):} Network SNR distribution}
    \label{fig5: SNR test dataset}
\end{figure}

To test the ability of \textsc{HYPERION} to recover $\vect{\theta}_{CE}$ and its overall performance, we have simulated an additional test set.  This set is composed of other $10^3$ simulated signals with the same distribution as the training one. 
The SNR distribution of the test set is shown in Fig. \ref{fig5: SNR test dataset}: both for the individual detectors and for the network. The network SNR shows, in particular, a peak around a value of 5 as seen in previous works \cite{Gleetter_Nunzio, detecting_and_PE_CE}. 

\textsc{HYPERION}'s inference has been compared with the one produced by \textsc{BILBY} \cite{BILBY}, adopting the \textsc{Dinesty} \cite{Dinesty} sampler. 
We tested different hyperparameters/settings, although with minimal discrepancies in the outputs. Henceforth we will refer to the results obtained with these settings: \text{r-walk} sampling method, \text{nlive $=1000$}, \text{nact $=50$}, \text{npool $=42$}.\\

\begin{center}
    \begin{figure}[!b]
        \centering
        \includegraphics[width = \textwidth]{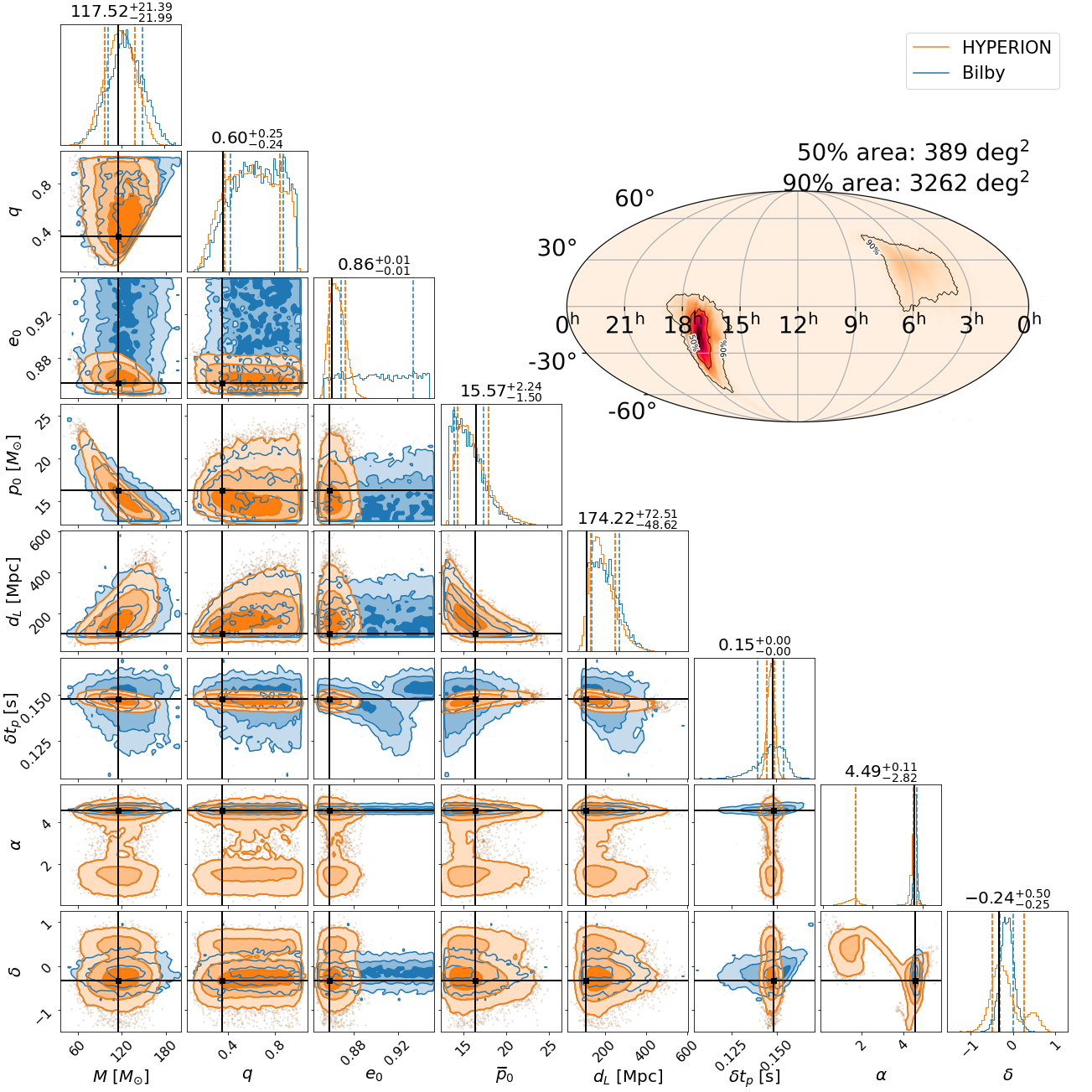}
        
        \caption{Comparison of posterior samples produced by \textsc{BILBY} and \textsc{HYPERION} for a test signal with network SNR $\simeq 30$ ($d_L \simeq 100$ Mpc). The posterior for most of the parameters are well overlapping, except for eccentricity $e_0$ which only \textsc{HYPERION} is able to estimate. On the other hand, \textsc{BILBY} gives a slightly better estimation of the localization.}
        \label{fig5: BILBY-HYPERION comparison 1}
    \end{figure}
\end{center}

\begin{center}
    \begin{figure}[!b]
        \centering
        \includegraphics[width = \textwidth]{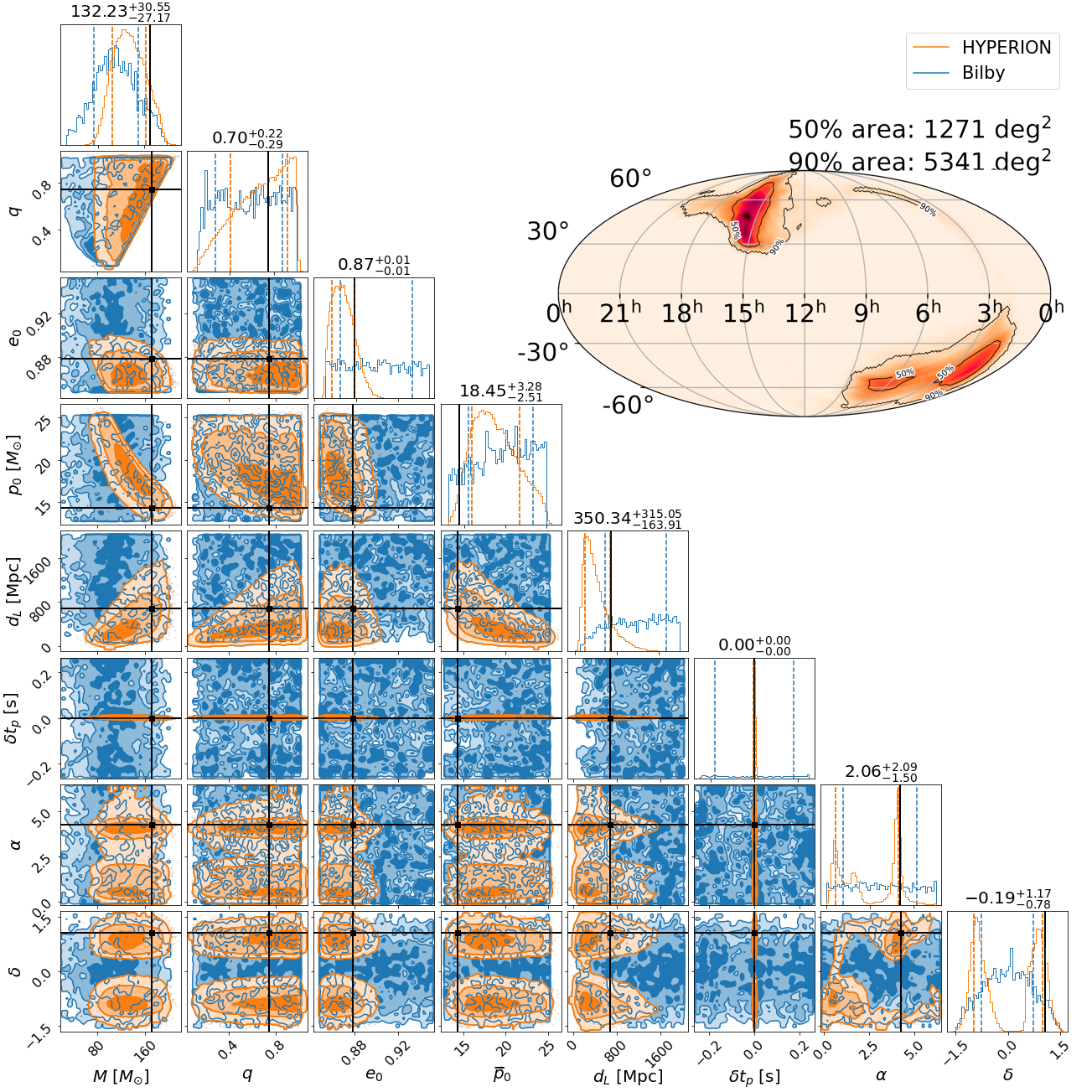}
        
        \caption{Comparison of posterior samples produced by \textsc{BILBY} and \textsc{HYPERION} for a test signal with network SNR $\simeq 12$ ($d_L \simeq 700$ Mpc). In this case, only \textsc{HYPERION}'s posteriors are informative since \textsc{BILBY} essentially reproduces the priors. $\delta t_p$ is the better-estimated parameter. The sky-localization's posteriors show bimodality: the dominant mode is, however, the one containing the right value for ($\alpha$, $\delta$).}
        \label{fig5: BILBY-HYPERION comparison 3}
    \end{figure}
\end{center}

\begin{center}
    
    \begin{figure}[!b]
        \centering
        \includegraphics[width = \textwidth]{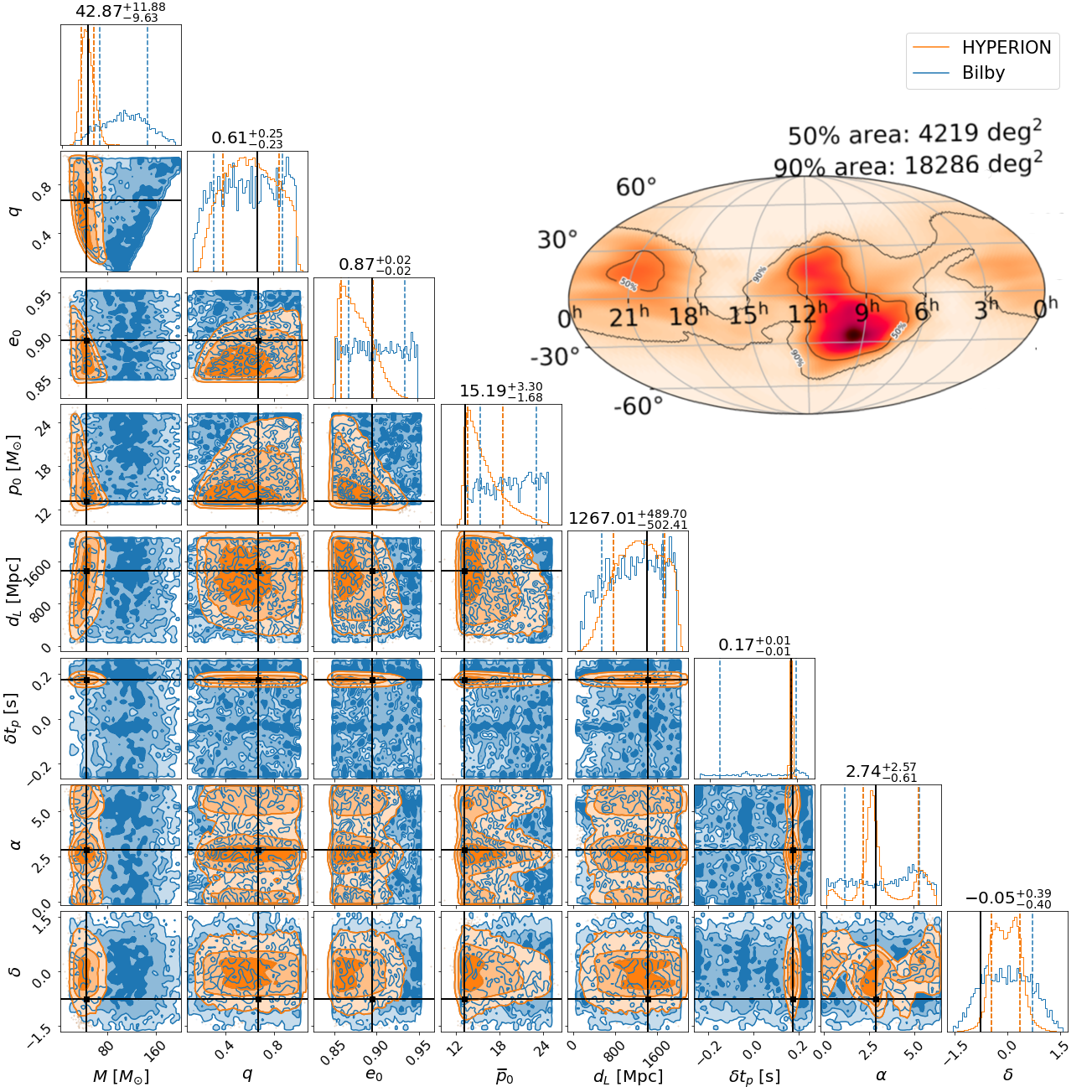}
        
        \caption{Comparison of posterior samples produced by \textsc{BILBY} and \textsc{HYPERION} for a test signal with network SNR $\simeq 6$ ($d_L \simeq 1400$ Mpc). In this case, only \textsc{HYPERION}'s posteriors are informative since \textsc{BILBY} reproduces the priors. The estimate of $M$ is emblematic as \textsc{BILBY} completely misses the right value, which is correctly estimated by \textsc{HYPERION}. The greater sky-localization area can be due to this signal peaking at lower frequencies where the sensitivity is worse. }
        \label{fig5: BILBY-HYPERION comparison 5}
    \end{figure}
\end{center}

With these parameters, $5\times 10^3$ posterior samples were obtained in $\sim$ 10 hours. Using the same hardware, we produced $5\times 10^4$ samples in 16 seconds by HYPERION running on the CPU only. When using the GPU, the same amount of samples were produced in just 0.5 seconds, improving by almost 5 orders of magnitude over standard Bayesian methods. 
Even on a CPU, the model can exploit, at most, hardware parallelization offered by the \textsc{PyTorch} deep learning library. The higher inference time required by \textsc{BILBY} ($\mathcal{O}(10\text{ h}$) is mainly due to the elevated number $\sim 10^7-10^8$ of likelihood evaluations required and the account for the autocorrelation time in the MCMC chains. 
In Figs. [\ref{fig5: BILBY-HYPERION comparison 1} $-$ \ref{fig5: BILBY-HYPERION comparison 5}] we show corner plots comparing the obtained posteriors for some of the simulated test signals. The upper quantiles, as well as the sky-maps, refer to \textsc{HYPERION}. The sky-maps, in particular, are produced with a subset of $10^4$ samples with the tool \texttt{ligo.skymap} \cite{ligo_skymap}.

\newpage

\newpage

\newpage

\section{Discussion}\label{Discussion}
The results obtained by testing HYPERION on simulated data show a very promising performance when compared with traditional parameter estimation based on Bayesian inference. The agreement between the parameter's values estimated by HYPERION and Bilby (e.g., in Fig. \ref{fig5: BILBY-HYPERION comparison 1}) shows that NFs are a viable and robust alternative to traditional Bayesian methods since they provide the same accuracy on results but on much shorter timescales.
At the same time, as shown in Fig. \ref{fig5: BILBY-HYPERION comparison 5}, HYPERION maintains the capability of providing informative posteriors even in the presence of low SNR signals. This can be seen, for instance, in the estimate of the total mass $M$ in Fig. \ref{fig5: BILBY-HYPERION comparison 5}, where HYPERION correctly produces values peaked around the simulated values 
(e.g., $M \simeq 40 \:\text{M}_\odot$) instead of reproducing the prior distribution. Results on posteriors using Bilby suggest that low SNR signals might require additional fine-tuning of the nested sampling hyperparameters. \\
We note that the time shift $\delta t_p$ parameter has narrow marginalized posteriors. This illustrates the efficiency of the Embedding Network and, in particular, of its Convolutional layers, which are able to recognize CE patterns even in the lowest SNR scenarios. In fact, one of the advantages of a time domain representation is that time-related patterns are directly accessible, in opposition to a frequency domain representation in which they manifest as phase shifts. 
Therefore, \textsc{HYPERION} is able to work as a standalone detection pipeline by using the 
Bayes factor statistic (Sec. \ref{subsec4: model selection with NFs Importance Sampling}).
When analyzing simulated data containing only noise, the posterior for $\delta t_p$ produced by HYPERION gets excessively broad, resembling the prior, thus indicating that the embedding network found no matches with known signals in the data.

As far as sky localization is concerned, we expect CE waveforms to be more difficult to localize than longer CBC waveforms, given their shorter duration and/or lower SNR. Indeed, with a shorter signal, it becomes more difficult to estimate the relative temporal shifts between the detectors because that information is concentrated in time. As a consequence, sky localization area increases with lower SNR or waveforms peaked at lower frequencies, as in Fig. \ref{fig5: BILBY-HYPERION comparison 5}. Although this aspect affects both standard methods and \textsc{HYPERION}, we notice that the latter provides better performance on localization for low SNR signals. This can be interpreted as proof of the efficiency of the localization CNN block in HYPERION's embedding network.

It can be further noticed that those posteriors (in particular the right ascension $\alpha$) show multimodality, which is related to periodicity in coordinates that induces a degeneracy for values near $0$ and $2\pi$. This multimodality is also a manifestation of the ability of NFs to model complicated distributions. 

\begin{figure}[!ht]
        \centering
        \includegraphics[width = \textwidth]{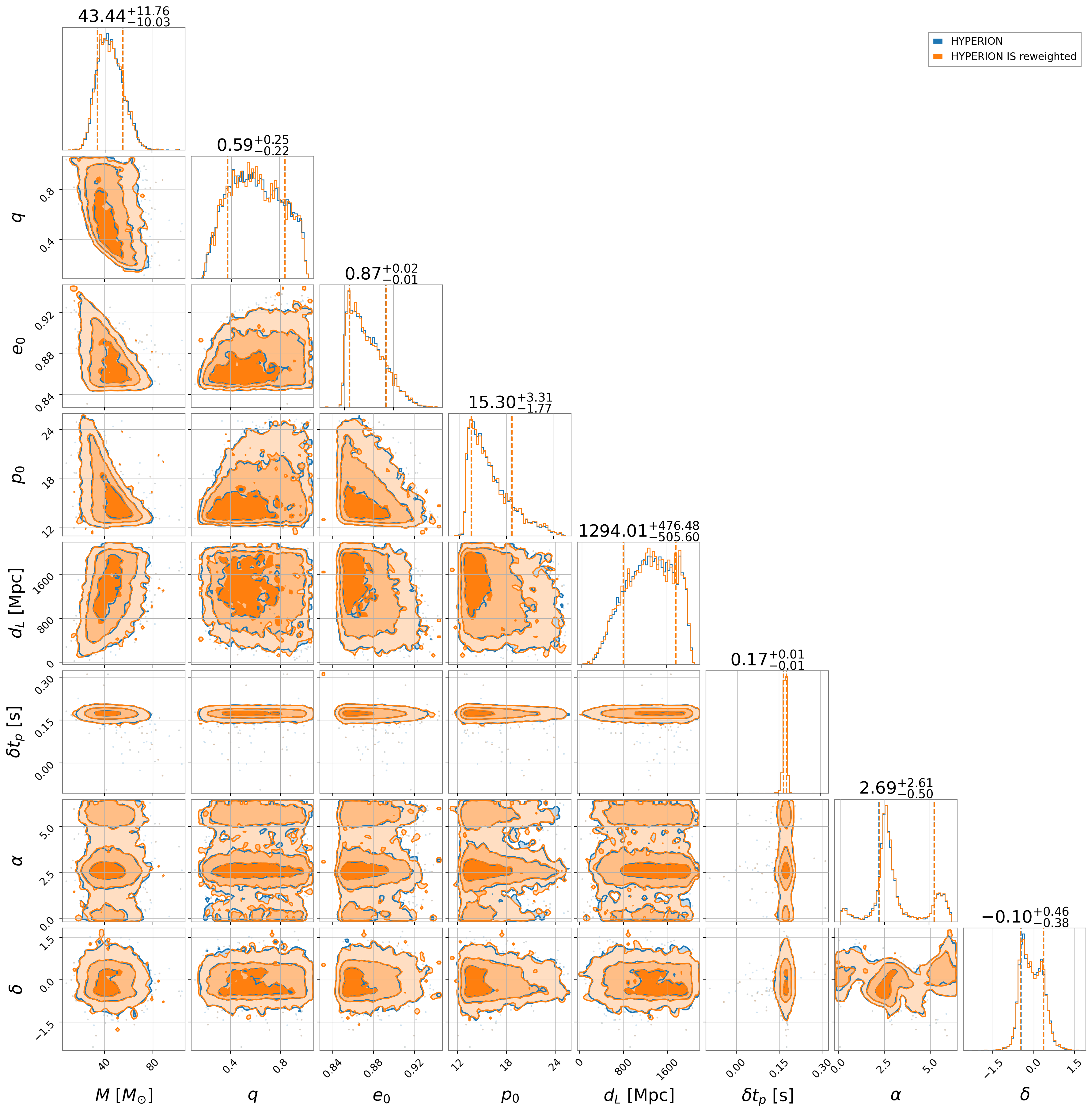}
        
        \caption{Comparison between the posterior inferred by HYPERION and the importance-reweighted posterior for the test sample of Fig. \ref{fig5: BILBY-HYPERION comparison 5}}
        \label{fig5: HYPERION-IS comparison}
    \end{figure}

To futher validate HYPERION's results we reweighted the posteriors with Importance Sampling, with the method described in Sec. \ref{subsec4: model selection with NFs Importance Sampling} and compared the two distributions. A metric for the inferred posterior's goodness can be defined as $\epsilon = \frac{1}{n}{(\sum_i w_i)^2}/({\sum_i w_i^2}) \in (0, 1]$ (\textit{sample efficiency)} \cite{DINGO_importance_sampling} where $w_i$ are the importance weights and $n$ the total number of posterior samples. 
We show an example in Fig. \ref{fig5: HYPERION-IS comparison} for which $\epsilon \approx 0.8$. We obtain similar results also for other test samples. The high efficiency can be justified by the fact that the test samples comes from the same distribution as the training ones (i.e. there are no OOD samples).

Although the results on the test set suggest the good performance of \textsc{HYPERION}, it is crucial to provide more accurate metrics to assess the power of this approach. Given the probabilistic and Bayesian nature of the model, a suitable test for its accuracy is the Probability$-$Probability Plot. This is a test used in Bayesian data analysis and widely adopted in the context of gravitational waves. 
The idea behind this test is to give a frequentist interpretation of Bayesian Credible Levels for the 1D marginalized posterior distributions. As an example, given a $CL=0.8$, for an optimal model, it means that in the $80\%$ of the cases, the true parameter value will lie in an interval that encloses $80\%$ of the posterior probability, regardless of the skewness of the distribution. \\
To perform the test, we first drew a set of $N$ data samples from the test set. For each of them we computed the posteriors and determined the percentile score of the true $\vect{\theta}_{CE}$ parameter values in each marginalized posterior. We then took the cumulative distribution (CDF) for each of the $\vect{\theta}_{CE}$. As the optimal case is represented by the percentiles being distributed according to a uniform one $\mathcal{U}(0, 1)$, we tested whether the CDFs lay on the diagonal. 

\begin{figure}[!t]
    \centering
    \includegraphics[width=\textwidth]{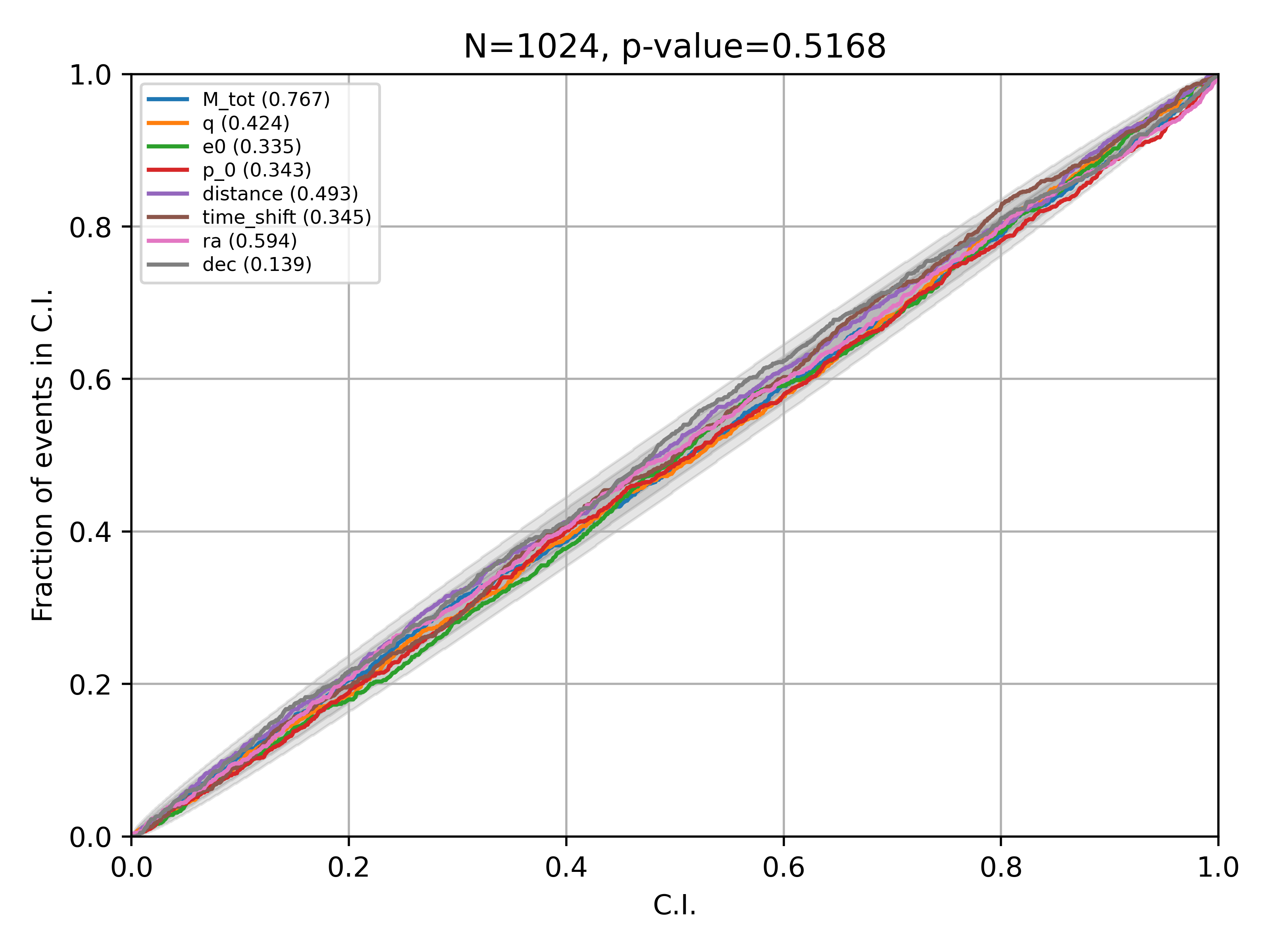}
    \caption{{Probability-Probability} Plot for a set of 1024 posterior evaluations from the Test Set. Each cumulative distribution lines up pretty well along the diagonal with a spread limited within the $2\sigma$ (grey regions) for almost all the CL interval. In the legend is also reported for each $\vect{\theta}_{CE}$ the KS statistics result. This plot has been made with a \textsc{Bilby} built-in function.}
    \label{fig5: PP plot}
\end{figure}

Fig. \ref{fig5: PP plot} shows the result of the PP test for a set of $N=1024$ draws. It is possible to notice that all the CDFs are well distributed along the diagonal with minimal spread limited within $2\sigma$ for almost all the CL intervals. To quantify how close the percentile distributions are to a uniform, a two-tail Kolmogorov-Smirnov Test is performed. The output $p-$values are shown in Fig. \ref{fig5: PP plot} as well. For each of the $\vect{\theta}_{CE}$, the $p-$values are greater than $0.1$ with the combined one $\simeq 0.5$, thus implying a good recovery of the parameters. Assuming a confidence level of $95\%$ (threshold at $\alpha = 0.05$), it is therefore not possible to reject the null hypothesis that the obtained CDFs are drawn from a uniform distribution since $p > \alpha$.
The parameter with the highest $p-$value is the total mass $M$ reaching $\simeq 0.7$, which does not surprise given the strong dependency of the Effective Fly-by waveforms on it. 

One of the main differences between a Normalizing Flow model such as \textsc{HYPERION} and standard methods is that it does not use Markov chains. MCMC algorithms have to account for correlation within the chains by thinning them. This has an impact on the efficiency since the the number of effective samples is reduced, or equivalently, to obtain the same $N_{\text{eff}}$ samples longer chains need to be produced (see Sec. 3 of \cite{Sokal_autocorrelation}). On the contrary, NFs are able to draw a set of $N$ independent samples directly, and to prove it we computed the autocorrelation time. In particular, $\hat{\tau}_\theta$ is determined for each of the $\vect{\theta}_{CE}$ set of samples with 
\begin{equation}\label{eq5: autocorrelation time}
    \hat{\tau}_\theta = 1 + 2 \sum_{\tau=1}^M \hat{c}_{\theta}(\tau) 
\end{equation}
where $\hat{c}(\tau)$ is the autocorrelation function computed with the Fast Fourier Transform algorithm, and $M$ is the first $\tau$ value for which the autocorrelation exceeds a threshold value ($\hat{c}_\theta(\tau) < 0.01$). 

Applying Eq.\eqref[]{eq5: autocorrelation time} the estimated autocorrelation time is $\hat{\tau}_\theta = 3\; \forall \theta$ : the smallest amount possible. The outlined procedure has been repeated for several different posteriors with no changes in the results. This hence indicates that \text{all} the posterior samples produced by \textsc{HYPERION} are valid. 

We address now the major limitations of this work and how they can be alleviated in the future. Being this work a proof of concept in the analysis of Close Encounters, we chose limited prior bounds for the simulations. However, extended simulations can be carried out anytime. The training dataset size can be, therefore, accordingly increased, provided that it is possible to account for the higher training time. Besides, our simulations assumed Gaussian and stationary noise. By considering also artifacts like nonstationarity and/or glitches in the simulations, this inference scheme can be made even more robust.

\section{Conclusions}\label{Conclusions}
In this work, we introduced HYPERION, a deep learning-based pipeline to detect and perform Bayesian parameter estimation on gravitational wave signals produced by binary close encounters. 

No firm detection of gravitational waves from close encounters has been achieved so far, making these sources particularly interesting to broaden our view of the gravitational wave Universe. Detecting and measuring parameters of close encounters could, therefore, help to shed light on the dynamical formation channels of compact binaries and explain the observed population. 
Furthermore, their detection would confirm the expectations of a sub-population of compact binaries merging with non-null eccentricities.\\
Moreover, their low-latency detection would allow the trigger of electromagnetic follow-up observations necessary to study a potential electromagnetic counterpart as well as the surrounding environment.
Detecting close encounters is difficult because of their intrinsic low signal-to-noise ratio, which makes them a hard target for current interferometers. Moreover, the short duration of the expected gravitational wave signal impacts the capability to estimate the sky coordinates and other parameters.
Deep learning is a promising tool for fast analysis of gravitational wave data that could constitute a viable approach for the study of this particular source. Since the standard methods for parameter estimation are based on a Bayesian framework, we explored the application of probabilistic machine learning. In particular, we focused on Normalizing Flows, an emerging machine-learning technique that is able to infer posterior distributions on very short timescales. Compared to other methods, such as MCMC, that require many likelihood computations, NFs introduce a faster posterior sampling based on a likelihood-free approach.

The architecture of HYPERION consists of two main parts: an embedding network whose goal is to extract features from the strain time series collected by a network of ground-based interferometers and an Affine Coupling Flow for the quick reconstruction of the posterior distribution and the estimation of the source parameters.
The training of HYPERION pipeline was carried out adopting the Effective Fly-by waveforms on a set of $\sim$5$\times 10^{6}$ simulated signals, obtaining extremely promising results on the test set. 
The value of the reconstructed parameters is consistent with the simulated values even in low signal-to-noise ratio cases. Furthermore, the HYPERION pipeline is $\sim$5 orders of magnitudes faster than traditional algorithms, providing the reconstruction of the posterior distribution on timescales of \SI{0.5}{\,s} instead of $\sim \SI{10}{\, h}$.\\
These results show that the NF-based approach is a viable and robust strategy for real-time detection and parameter estimation of signals from close encounters, also enabling electromagnetic follow-up campaigns.\\
There are several other prospects about how this work might be extended or improved in the future. In this work, we focused in particular on CE signals from binary black holes as they are the most likely to be observed with the current generation detector, both in terms of SNR and expected rates. Nevertheless, CE emission is expected also from systems containing neutron stars, and they constitute a potential source for ground-based third-generation detectors like Einstein Telescope or Cosmic Explorer \cite{Punturo_ET,evans_ce},  or spaceborne missions like LISA \cite{LISA23}. 
Future work will include the analysis of repeated bursts from multiple periastron encounters, allowing to track the evolution of orbital parameters during the inspiral phase. 

The deep learning method presented in this work will permit rapid systematic searches for transients produced by close encounters, with the exciting possibility of detecting these signals and exploring the formation scenarios of binary compact systems in the Universe.
Furthermore, this inference scheme is not limited to gravitational waves emitted by close encounters. With minimal changes, e.g., by employing a different waveform model during training and/or changing its hyperparameters, HYPERION can be adapted to search for other kinds of sources, e.g., other kinds of burst-like signals.

\begin{acknowledgments}
The authors gratefully acknowledge NVIDIA Corporation for donating the two A30 GPU used in this work. 
L.P. acknowledges the support of the PhD scholarship in Physics "High Performance Computing and Innovative Data Analysis Methods in Science" (Cycle XXXVIII, Ministerial Decree no. 351/2022) and received funding from the European Union Next-Generation EU - National Recovery and Resilience Plan (NRRP) – MISSION 4 COMPONENT 1, INVESTMENT N.4.1 – CUP N.I51J22000630007. This manuscript reflects only the authors’ views and opinions, neither the European Union nor the European Commission can be considered responsible for them. 

\end{acknowledgments}

\bibliography{references}

\end{document}